\documentclass[aps,prc,preprint,floatfix,showpacs,amsmath,amssymb]{revtex4}
\usepackage{graphicx}
\usepackage{amssymb}
\usepackage{dcolumn}
\usepackage{color} 
\begin{document}
\begin{flushright}
SLAC-PUB-13582\\
\end{flushright}
\title{Hamiltonian light-front field theory in a basis function approach}

\author{J.~P.~Vary$^{1}$, H.~Honkanen$^{1}$, Jun~Li$^{1}$, P.~Maris$^{1}$,
S.~J.~Brodsky$^{2}$, A.~Harindranath$^3$, G.~F.~de~Teramond$^4$, 
P.~Sternberg$^{5}$\footnote{currently at ILOG Inc, Incline Village, NV}, E.~G.~Ng$^{5}$, C.~Yang$^{5}$}

\affiliation{
$^1$Department of Physics and Astronomy, Iowa State University, Ames, Iowa 50011, USA \\
$^2$SLAC National Accelerator Laboratory, Stanford University, Menlo Park, California, USA \\
$^3$Theory Group, Saha Institute of Nuclear Physics,1/AF, Bidhannagar, Kolkata, 700064, India\\
$^4$Universidad de Costa Rica, San Jos\'e, Costa Rica\\
$^5$Lawrence Berkeley National Laboratory, Berkeley, California, USA}

\date{\today}

\begin{abstract}
Hamiltonian light-front quantum field theory 
constitutes a framework for the non-perturbative solution of invariant masses and correlated 
parton amplitudes of self-bound systems. By choosing the light-front gauge and adopting
a basis function representation, we obtain a large, sparse, Hamiltonian matrix
for mass eigenstates of gauge theories that is solvable by adapting the {\it ab initio} no-core methods 
of nuclear many-body theory. Full covariance is recovered in the continuum limit, the infinite matrix limit. There is considerable freedom in the choice of the orthonormal and complete set of
basis functions with convenience and convergence rates providing key considerations.
Here, we use a two-dimensional harmonic oscillator basis for transverse
modes that corresponds with eigensolutions of the soft-wall
AdS/QCD model obtained from light-front holography.  We outline our approach and present 
illustrative features of some non-interacting systems in a cavity.  We illustrate the first steps towards
solving QED by obtaining the mass eigenstates of an electron in a cavity in small basis spaces and discuss the computational challenges.

\end{abstract}
\pacs{11.10.Ef,11.15.Tk}
\maketitle
%

\section{Introduction}
Non-perturbative Hamiltonian light-front quantum field theory presents opportunities 
and challenges that bridge particle physics and nuclear physics.
Major goals include predicting both the masses and transitions rates of the
hadrons and their structures as seen in high-momentum transfer experiments.
Current focii of intense experimental and theoretical research that could 
benefit from insights derived within this Hamiltonian approach 
include the spin structure of the proton, the neutron electromagnetic form factor,
the generalized parton distributions of the baryons, etc.

Hamiltonian light-front field theory 
in a discretized momentum basis \cite{BPP98} and in transverse lattice
approaches \cite{BD2002,JPV214,GP08} have shown significant promise.
We present here a basis-function approach that exploits
recent advances in solving the non-relativistic strongly interacting nuclear
many-body problem \cite{NCSM, NCFC}.  We note that both light-front field theory and
nuclear many-body theory face common issues within the Hamiltonian approach - i.e. how to 
(1) define the Hamiltonian;
(2) renormalize to a finite space;
(3) solve for non-perturbative observables while preserving as many symmetries as possible; and,
(4) take the continuum limit.
In spite of the technical hurdles, Ken Wilson has assessed the advantages of adopting advances in quantum many-body theory and has long advocated adoption of basis function methods as an alternative to the lattice gauge approach \cite{KW1989}.

There are three main advantages of Hamiltonian light-front quantum field theory
motivating our efforts to overcome the technical hurdles.  First, one evaluates experimental observables that are non-perturbative and relativistically invariant quantities such as masses, form factors, structure functions, etc.  Second, one evaluates these quantities in Minkowski space and, third, there is no fermion doubling problem.

We begin with a brief overview of recent advances in solving 
light nuclei with realistic nucleon-nucleon (NN) and three-nucleon (NNN)
interactions using {\it ab initio} no-core methods in a basis function representation. 
Then, we introduce our basis function approach to
light-front QCD within the light-front gauge.  
Renormalization/regularization issues are also addressed.
We present illustrative features of our approach with the example
of cavity-mode QED and sketch its extension to cavity-mode QCD.
For a specific QED example, we work in small basis spaces and solve for 
the mass eigenstates of an electron coupled to a single photon in a transverse
harmonic oscillator cavity. 

The present work is an expanded version of a recent paper where we provided an initial
introduction to our approach \cite{LC2008}.

\section{No Core Shell Model (NCSM) and No Core Full Configuration (NCFC) methods}

To solve for the properties of nuclei, self-bound strongly interacting systems, with
realistic Hamiltonians, one faces immense analytical and computational
challenges.  Recently, {\it ab initio} methods have been developed that
preserve all the underlying symmetries and 
converge to the exact result.  The basis function approach that we adopt here
\cite{NCSM, NCFC} is one of several methods shown to be successful.  
The primary advantages are its flexibility for choosing the Hamiltonian, the method
of renormalization/regularization and the basis space.  
These advantages impel us to adopt the basis function approach 
in light-front quantum field theory.  While non-relativistic applications in finite nuclei restrict the basis to a fixed number of fermions, we introduce here the extension to a flexible number of fermions,
antifermions and bosons.

Refs. \cite{NCSM} and \cite{NCFC} provide examples of the recent advances in the {\it ab initio} NCSM and NCFC, respectively.  The former adopts 
a finite basis-space renormalization method and applies it to realistic nucleon-nucleon (NN) and three-nucleon (NNN) interactions (derived from chiral effective field theory) to solve nuclei with Atomic Numbers $A =10-13$ \cite{chiral07}. Experimental binding energies, spectra, electromagnetic moments and transition rates are well-reproduced.
The latter adopts a realistic NN interaction that is sufficiently soft  that renormalization is not necessary and binding energies obtained from a sequence of finite matrix solutions may be extrapolated to the infinite matrix limit.   Owing to uniform convergence and the variational principle, one is also able to assess the theoretical uncertainties in the extrapolated result.  One again obtains good agreement with experiment. 

It is important to note the analytical and technical advances made to solve these problems. First, non-perturbative renormalization has been developed to accompany these basis-space methods that preserve all the symmetries of the underlying Hamiltonian including highly precise treatments of the center-of-mass motion.  Several schemes have emerged with impressive successes and current research focuses on
detailed understanding of the scheme-dependence of convergence rates (different observables converge at different rates) \cite{Bogner07}. Second, large scale calculations are performed on leadership-class parallel computers, at Argonne National Laboratory and at Oak Ridge National Laboratory, to solve for the low-lying eigenstates and eigenvectors as well as to carry out evaluation of a suite of experimental observables.  For example, one can now obtain the low-lying solutions for $A = 14$ systems with matrices of dimension one to three billion on 8000 to 50000 processors within a few hours of wallclock time.  Since the techniques are evolving rapidly \cite{Sternberg08} and the computers are growing dramatically, much larger matrices are within reach.

In a NCSM or NCFC application, one adopts a 3-D harmonic oscillator for all the particles in the nucleus (with harmonic oscillator energy $\Omega$), treats the neutrons and protons independently, and generates a many-fermion basis space that includes the lowest oscillator configurations as well as all those generated by allowing up to $N_{max}$ oscillator quanta of excitations.  The single-particle states are formed by coupling the orbital angular momentum to the spin forming the total angular momentum $j$ and total angular momentum projection $m_j$.   The many-fermion basis consists of states where particles occupy the allowed orbits subject to the additional constraint that the total angular momentum projection $M_j$ is a pre-selected value. This is referred to as the $m$-scheme basis and, in a single run, one obtains eigenstates with total angular momentum $J \ge M_j$. For the NCSM one also selects a renormalization scheme linked to the many-body basis space truncation while in the NCFC the renormalization is either absent or of a type that retains the infinite matrix problem.  In the NCFC case \cite{NCFC}, one extrapolates to the continuum limit as we now illustrate.

\begin{figure}[tb]
\includegraphics[width=0.7\columnwidth]{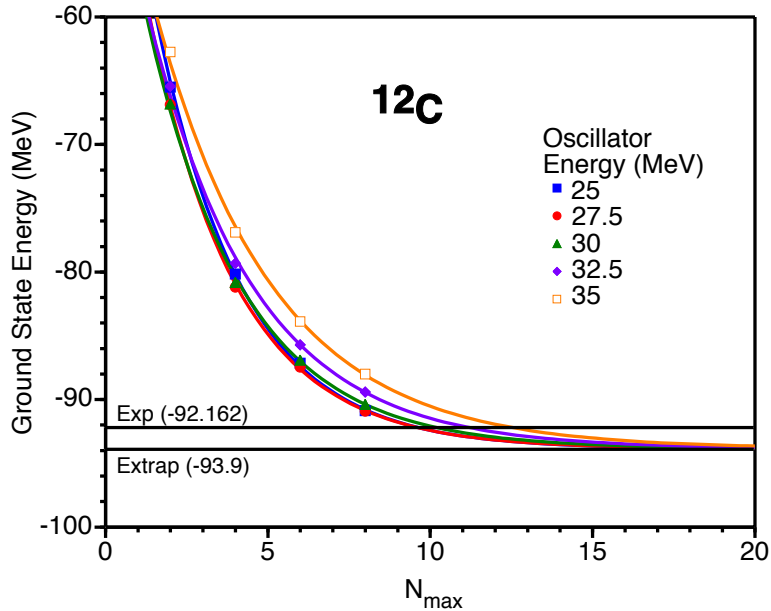}
\caption{(color online) Calculated ground state energy of $^{12}$C for
$N_{max}=2{-}8$ (discrete points) at selected values of the oscillator energy, $\Omega$.  
For each $\Omega$, the results are fit to an exponential plus
a constant, the asymptote, which is constrained to be the same for 
each curve\cite{NCFC}.  We display the experimental
ground state energy and the common asymptote.} 
\label{12C}
\end{figure}

We show in Fig. \ref{12C} results for the ground state of $^{12}C$ as a function of $N_{max}$
obtained with a realistic NN interaction, JISP16 \cite{Shirokov07}.  The smooth curves portray fits that achieve the desired independence of  $N_{max}$ and $\Omega$ so as to yield the extrapolated ground state energy.   Our assessed uncertainty in the extrapolant is about 2 MeV and there is rather good agreement with experiment within that uncertainty.  The largest cases presented in Fig. \ref{12C} correspond to $N_{max}=8$, where the matrix reaches a basis dimension near 600 million.  $N_{max}=10$ produces a matrix near 8 billion and its lowest eigenvalues have now been solved at two values of  $\Omega$.  These $N_{max}=10$  results follow closely the curves shown in Fig. \ref{12C}  and will be presented elsewhere.

\section{Choice of Representation for Light Front Hamiltonians}

It has long been known that light-front Hamiltonian quantum field theory has similarities with non-relativistic quantum many-body theory.  We further exploit this connection, in what we will term a ``Basis Light Front Quantized (BLFQ)" approach, by adopting a light-front single-particle basis space consisting of the 2-D harmonic oscillator for the transverse modes (radial coordinate $\rho$ and polar angle $\phi$) and a discretized momentum space basis for the longitudinal modes.  Adoption of this basis is also consistent with recent developments in AdS/CFT correspondence with QCD \cite{Karch:2006pv,Erlich:2005qh,deTeramond:2008ht,Brodsky:2003px,Polchinski:2001tt}.  In the present application to the non-interacting problem, we will adopt periodic boundary conditions (PBC) for the longitudinal modes and we omit the zero mode. For our light-front coordinates, we define $x^{\pm}=x^0 \pm x^3$, $x^{\perp}=(x^1,x^2)$ and coordinate pair $(\rho,\phi)$ are the usual cylindrical coordinates in $(x^1,x^2)$.  The variable $x^+$ is light-front time and $x^-$ is the longitudinal coordinate.  We adopt $x^+=0$, the ``null plane", for our quantization surface.

The 2-D oscillator states are characterized by their principal quantum number $n$, orbital quantum number $m$ and harmonic oscillator energy $ \Omega $.   It is also convenient to interpret the 2-D oscillator as a function of the dimensionless radial variable  $\sqrt{M_0\Omega}\rho$ where $M_0$ has units of mass and $\rho$ is the conventional radial variable in units of length. Thus, the length scale for transverse modes is set by the chosen value of $\sqrt{M_0\Omega}$. 

The properly orthonormalized wavefunctions, $\Phi_{n,m}(\rho, \phi) = \langle \rho \phi | n m \rangle \ = 
f_{n,m}(\rho) \chi_{m}(\phi)$, are given in terms of the Generalized Laguerre Polynomials, $L_n^{|m|}(M_0\,\Omega\,\rho^2)$, by
\begin{eqnarray}
  f_{n,m}(\rho) &=& \sqrt{2\,M_0\,\Omega} \; \sqrt{\frac{n!}{(n+|m|)!}}\; 
   {\rm e}^{-M_0 \, \Omega \, \rho^2 / 2} \;
   \left(\sqrt{M_0\,\Omega}\;\rho\right)^{|m|} \;
   L_n^{|m|}(M_0\,\Omega\,\rho^2)
\label{Eq:wfn2dHOfx}
\\
  \chi_{m}(\phi) &=& \frac{1}{\sqrt{2\pi}} \, {\rm e}^{i\,m\,\phi}
\label{Eq:wfn2dHOchix}
\end{eqnarray}
with eigenvalues $E_{n,m}=(2n+|m|+1)\Omega$.
The orthonormalization is fixed by
\begin{eqnarray}
 \langle n m | n' m' \rangle \;=\;
  \int_0^\infty \int_0^{2\pi}\!\! \rho\,d\rho \; d\phi \; 
  \Phi_{n,m}(\rho, \phi)^* \; \Phi_{n',m'}(\rho, \phi) 
   &=&  \delta_{n,n'} \; \delta_{m,m'} 
\label{Eq:orthonormal}
\end{eqnarray}
which allows for an arbitrary phase factor ${\rm e}^{i\alpha}$ that we have taken to be unity.  One of the significant advantages of the 2-D oscillator basis is the relative ease for transforming results between coordinate space and momentum space.  That is, the Fourier transformed wavefunctions have the same analytic structure in both coordinate and momentum space, a feature reminiscent of a plane-wave basis.

\begin{figure}[tb]
\includegraphics[width=1.0\columnwidth]{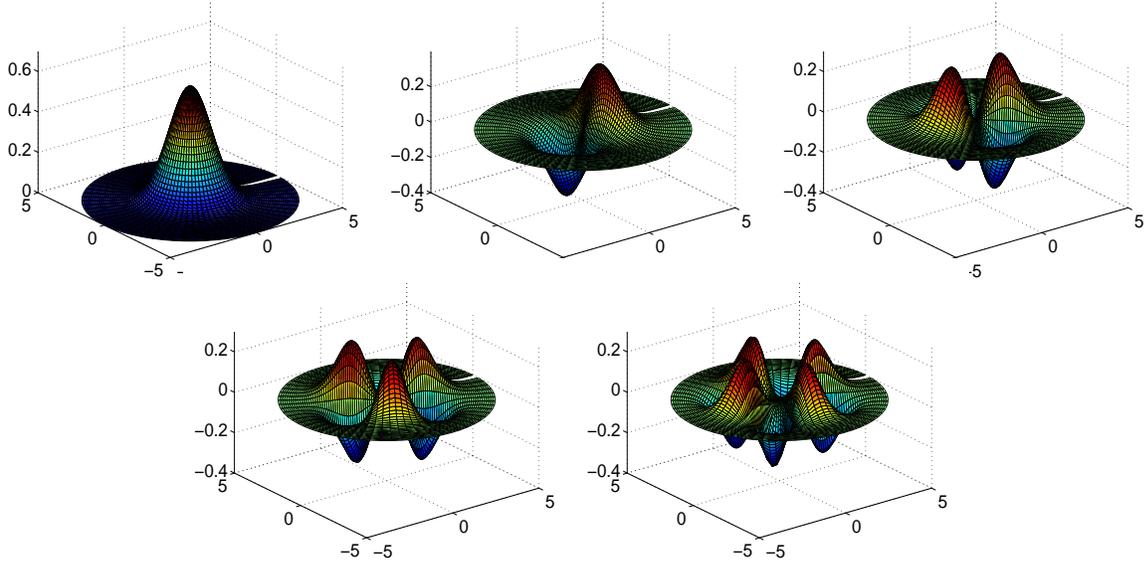}
\caption{(color online) Modes for $n=0$ of the 2-D harmonic oscillator selected for
the transverse basis functions. The orbital quantum number $m$ progresses across the rows by 
integer steps from 0 in the upper left to 4 in the lower right and counts the pairs of
angular lobes. Amplitudes as well as x-axis and y-axis coordinates are in dimensionless units.} 
\label{Neq0_modes}
\end{figure}

\begin{figure}[tb]
\includegraphics[width=1.0\columnwidth]{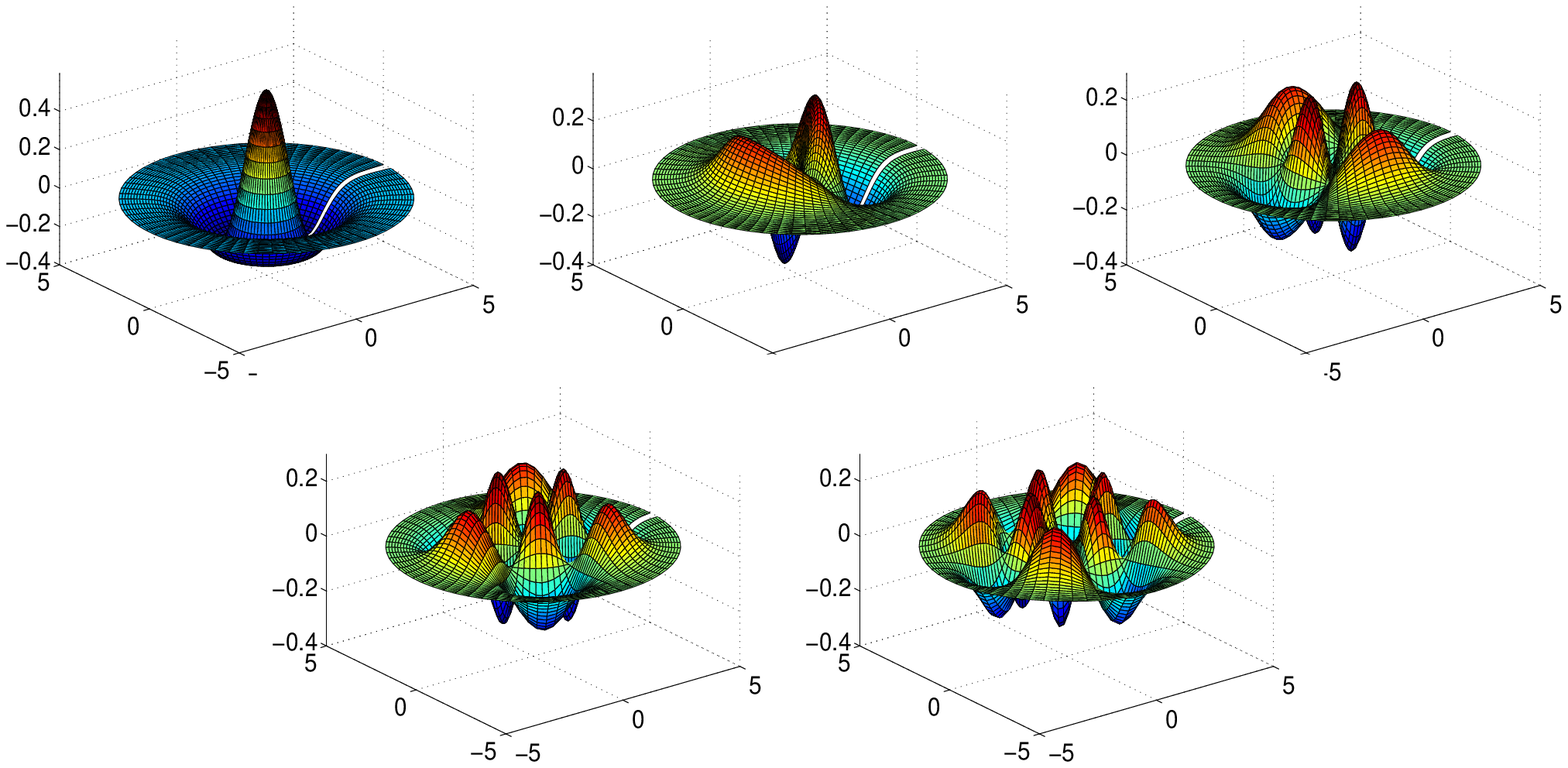}
\caption{(color online) Modes for $n=1$ of the 2-D harmonic oscillator selected for
the transverse basis functions. The orbital quantum number $m$ progresses across the rows by 
integer steps from 0 in the upper left to 4 in the lower right and counts the pairs of
angular lobes. Amplitudes as well as x-axis and y-axis coordinates are in dimensionless units.} 
\label{Neq1_modes}
\end{figure}

In order to gain an impression of the transverse modes in our light-front basis, we present in Figs. \ref{Neq0_modes} and \ref{Neq1_modes} snapshots of selected low-lying modes.  As one increases the orbital quantum number $m$,  pairs of maxima and minima populate the angular dependence of the basis function.  Also, as one increases the principal quantum number $n$, additional radial nodes appear as evident in the progression from Fig. \ref{Neq0_modes} to Fig. \ref{Neq1_modes}.

To provide a perspective on the full 3-D basis, we introduce longitudinal modes $\psi_{j}$ defined on $-L \le x^- \le L$ with both periodic boundary conditions (PBC) and antiperiodic boundary conditions (APBC).  We also introduce purely real form to be used in a figure below:

\begin{eqnarray}
  \psi_{k}(x^-) &=& \frac{1}{\sqrt{2L}} \, {\rm e}^{i\,\frac{\pi}{L}k\,x^-}
\label{Eq:longitudinal1}
\\
\psi_{k}(x^-) &=&\frac{1}{\sqrt{\pi L}} \sin{\frac{\pi}{L}k\,x^-}
\label{Eq:longitudinal2}
\end{eqnarray}
where $k=1,2,3,...$  for PBC (neglecting the zero mode) and $k=\frac{1}{2},\frac{3}{2},\frac{5}{2},...$ in Eq. \ref{Eq:longitudinal1} for APBC.  Similarly, $k=1,2,3$ in Eq. \ref{Eq:longitudinal2} for reflection antisymmetric amplitudes with box boundary conditions (amplitude vanishes at $x^- = \pm L$). The full 3-D single particle basis state is defined by the product form

\begin{eqnarray}
  \Psi_{k,n,m}(x^-,\rho,\phi) &=& \psi_{k}(x^-) \Phi_{n,m}(\rho, \phi).
\label{Eq:totalspwfn}
\end{eqnarray}

\begin{figure}[tb]
\includegraphics[width=0.9\columnwidth]{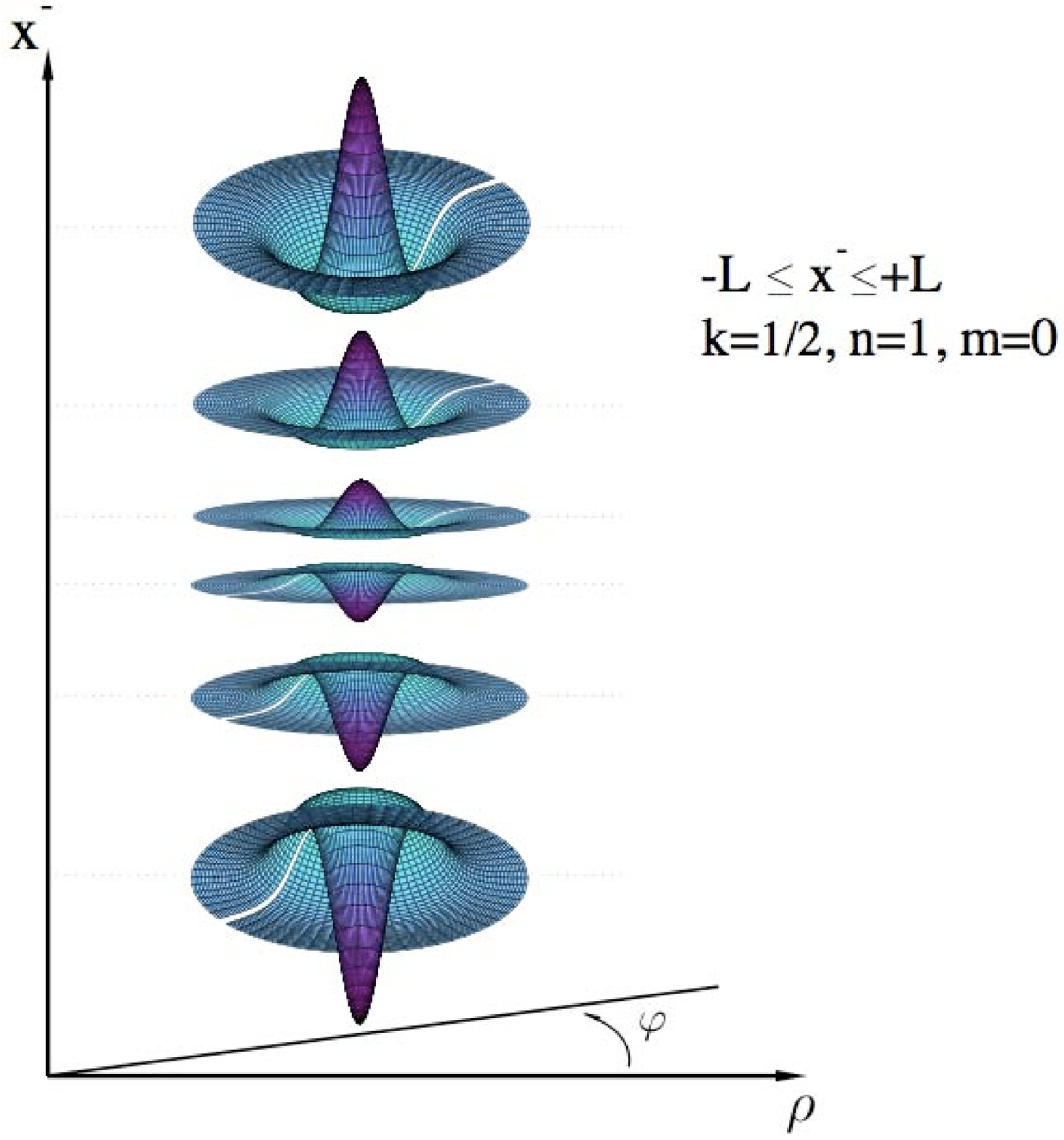}
\caption{(color online) Transverse sections of the real part of a 3-D basis function involving a 2-D harmonic oscillator and a longitudinal mode of Eq. \ref {Eq:longitudinal1} with antiperiodic boundary conditions (APBC). The quantum numbers for this basis function are given in the caption. The basis function is shown for the 
full range $-L \le x^- \le L$.} 
\label{APBC_3D_mode}
\end{figure}

\begin{figure}[tb]
\includegraphics[width=0.9\columnwidth]{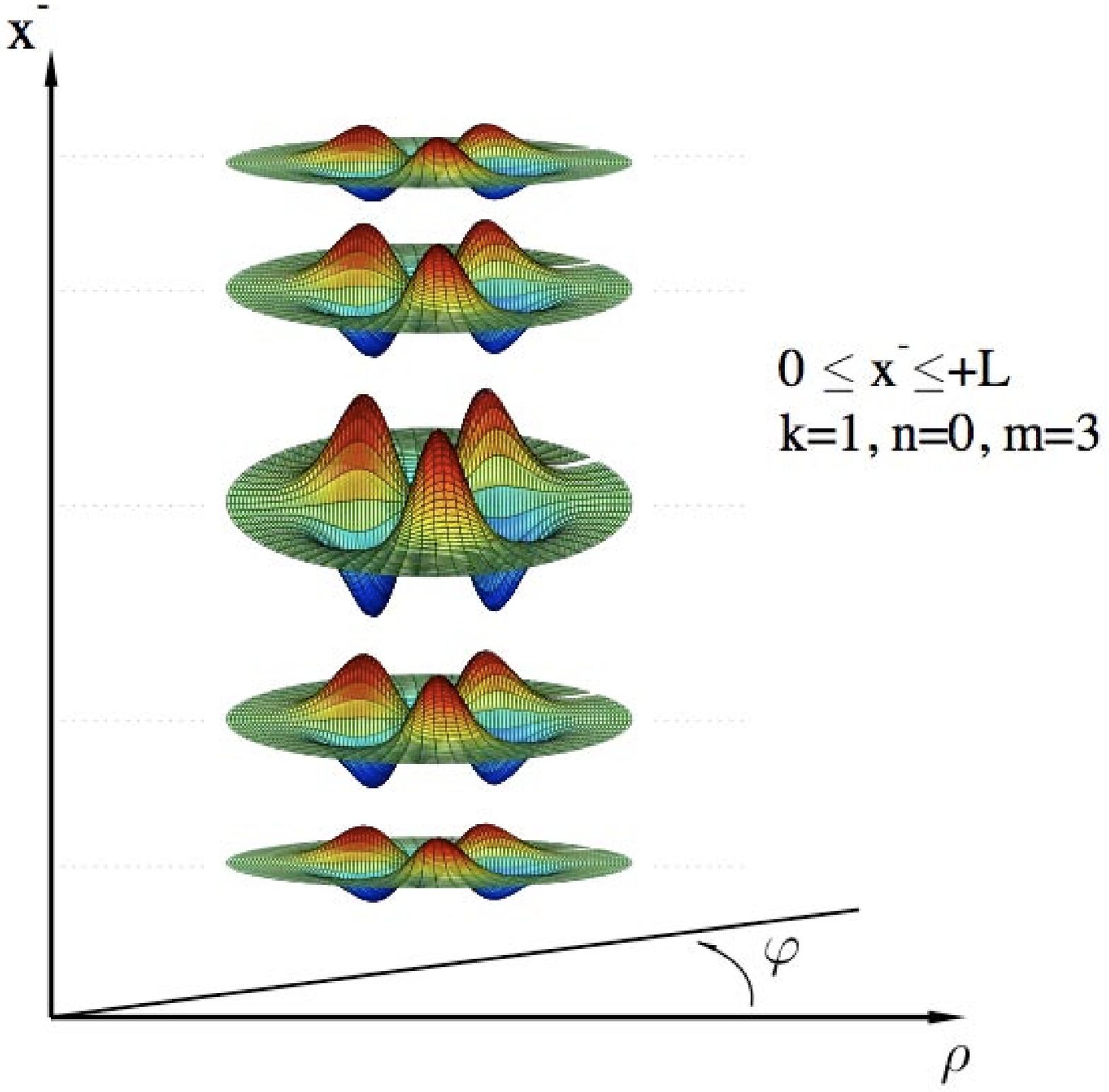}
\caption{(color online) Transverse sections of a 3-D basis function involving a 2-D harmonic oscillator and a longitudinal mode of Eq. \ref {Eq:longitudinal2} with box boundary conditions (wavefunction vanishes at $\pm$L). The quantum numbers for this basis function are given in the caption. The basis function is shown for positive values of $x^-$ and is antisymmetric with respect to $x^-=0$} 
\label{BoxBC_3D_mode}
\end{figure}

For a first illustration, we select a transverse mode with $n=1, m=0$ joined together with the $k=\frac{1}{2}$ longitudinal APBC mode of Eq. \ref{Eq:longitudinal1} and display slices of the real part of this 3-D basis function at selected longitudinal coordinates, $x^-$ in Fig. \ref{APBC_3D_mode}.  For comparison, we present a second example with Eq. \ref{Eq:longitudinal2} for the longitudinal mode in Fig. \ref{BoxBC_3D_mode}.  Our purpose in presenting both Figs. \ref{APBC_3D_mode} and \ref{BoxBC_3D_mode} is to suggest the richness, flexibility and economy of texture available for solutions in a basis  function approach.  Note that the choice of basis functions is rather arbitrary, including which boundary conditions are imposed, except for the standard conditions of orthonormality and completeness within the selected symmetries. 

Although our choice of basis functions is not dictated by theory, it is buttressed by the phenomenological success of the ``soft-wall"  AdS/QCD model~\cite{Karch:2006pv,Erlich:2005qh} which uses  a harmonic oscillator potential in the fifth dimension of Anti-de Sitter space to simulate color confinement.
As shown in ref. \cite{deTeramond:2008ht} one  can use  ``light-front holography"~\cite{Brodsky:2003px} to transform the bound state equations for the wavefunction in AdS space~\cite{Polchinski:2001tt} to  a corresponding bound-state equation in physical space at fixed light-front time $\tau$. The resulting light-front equation is similar in form 
to the  Schr\"odinger radial wave equation at fixed $t$ which describes the quantum-mechanical
structure of atomic systems. However, the formalism at fixed light-front time is relativistic and frame independent. Thus, for the specific example of a $q \bar q$ pair, one obtains a relativistic wave equation applicable to hadron physics, where the light-front coordinate $\zeta =b_\perp \sqrt{x(1-x)}$ plays the role of the radial variable $r$ of the nonrelativistic theory.   Here, $x$ is the light-front momentum fraction of the quark and $b_\perp$ is the magnitude of the transverse relative separation coordinate.
In this example, the meson eigenvalue equation is~\cite{deTeramond:2008ht} 
\begin{equation}
\left[-{d^2\over d\zeta^2} - {1-4 L^2\over 4 \zeta^2}+U(\zeta) \right]\phi(\zeta) = 
M^2\phi(\zeta),
\label{eq:AdSLFWF}
\end{equation}
where the complexity of the QCD interactions among constituents is summed up in the addition of the effective
potential $U(\zeta)$, which is then modeled to enforce confinement.  The potential in the soft wall model  is  $ U(\zeta) = \kappa^4 \zeta^2 + 2 \kappa^2(J-1)$ where $J$ is the total angular momentum of the hadron.  Using the substitution $ \phi(\zeta) = \zeta^{1/2} R(\zeta) $, $\kappa \zeta = \sqrt{M_0 \Omega} \rho $ and 
$L = |m|$, we arrive at the transverse 2-D harmonic oscillator wave equation whose solution is given in Eq. \ref{Eq:wfn2dHOfx}. 

There is one additional distinction between our choice of transverse basis functions and the solutions of the AdS/QCD model: we adopt single-parton coordinates as the basis function arguments while AdS/QCD adopts a relative coordinate between the constituents.  Our selection is natural for the applications within an external cavity that we present here and is most convenient for enforcing the boson and fermion statistics when dealing with arbitrary many partons.  In future work without the external cavity, we may invoke a Lagrange multiplier method, analogous to the method in the NCSM and NCFC approaches \cite{NCSM,NCFC}, to separate the relative motion from the total system's motion in the transverse direction.

The  solutions of the light-front equation (\ref{eq:AdSLFWF}) determine the masses of the hadrons,  given the total internal spin $S$, the orbital angular momenta $L$ of the constituents, and the index $n$,  the number of nodes of the wavefunction in $\zeta$.   For example, if the total quark spin $S$ is zero,  the meson bound state spectrum follows the quadratic form $M^2 = 4 \kappa^2 (n + L). $   Thus the internal orbital angular momentum $L$ and its effect on quark kinetic energy play an explicit role.  The corresponding wavefunctions of  the mesonic eigensolutions describe the probability distribution of the $q \bar q$ constituents
for the different orbital and radial states.  
The separation of the constituent quark and antiquark in AdS space get larger as the orbital angular momentum increases. 
The pion with $n=0$ and $L=0$ is massless for zero quark mass, in agreement with general arguments based on chiral symmetry. If the total spin of the constituents is  $S =1$, the corresponding mass formula for the orbital and radial spectrum of 
the $\rho$ and $\omega$ vector mesons is $M^2 = 4 \kappa^2 (n + L+ 1/2)$. The states are aligned along linear Regge trajectories and agree well with experiment.  The resulting  light-front wavefunctions also give a good account of the hadron form factors.

The AdS/QCD model, together with light-front holography,  provides  a semiclassical first approximation to strongly coupled QCD. The BLFQ approach in this paper provides a natural extension of the AdS/QCD light-front wavefunctions to multiquark and multi-gluonic Fock states, thus allowing 
for particle creation and absorption. By setting up and diagonalizing the light-front QCD Hamiltonian on this basis, 
we incorporate higher order corrections corresponding to the full QCD theory,
and we hope to gain insights into the success of the AdS/QCD model.

\section{Cavity mode light-front field theory without interactions}

For a first application of the BLFQ approach, we consider a non-interacting QED system confined to a transverse harmonic trap or cavity.   For simplicity, we take the spin $1/2$ leptons as massless. The basis functions are matched to the trap so we implement a transverse 2-D harmonic oscillator basis with length scale fixed by the trap and finite modes in the longitudinal direction with APBC.  
Since we are ultimately interested in the self-bound states of the system, we anticipate
adoption of the NCSM method for factorizing the eigensolutions into simple products of intrinsic and
total momentum solutions in the transverse direction \cite{NCSM}.  That is, with a suitable transverse
momentum constraint such as a large positive Lagrange multiplier times the 2-D harmonic oscillator Hamiltonian acting on the total transverse coordinates, the low-lying physical solutions will all have the same expectation value of total transverse momentum squared.  Therefore, following Ref. \cite{BPP98} we introduce the total invariant mass-squared $M^2$ for these low-lying physical states in terms of a Hamiltonian $H$ times a dimensionless integer for the total light front momentum $K$

\begin{eqnarray}
M^2 + P_{\perp}P_{\perp} \rightarrow  M^2 + const = P^+P^- = KH
\label{Mass-squared}
\end{eqnarray}
where we absorb the constant into $M^2$.  
The Hamiltonian $H$ for this system is defined by the sum of the occupied modes $i$ in each many-parton state with the scale set by the combined constant $\Lambda^2 = 2M_0\Omega$:
\begin{eqnarray}
H = 2M_0 P^-_c = \frac{2M_0\Omega}{K}\sum_i{\frac{2n_i+|m_i| +1}{x_i}}.
\label{Hamiltonian}
\end{eqnarray}

We adopt symmetry constraints and two cutoffs for our many-parton states.  For symmetries, we fix the total charge $Z$, the total azimuthal quantum number $M_t$, and the total spin projection $S$ along the $x^-$ direction.  For cutoffs, we select the total light-front momentum, $K$, and the maximum total quanta allowed in the transverse mode of each many-parton state, $N_{max}$.  For the longitudinal modes, we select those with PBC from Eq. \ref{Eq:longitudinal1}. The chosen symmetries and cutoffs are expressed in terms of sums over the quantum numbers of the single-parton degrees of freedom contained in each many-parton state of the system in the following way:

\begin{eqnarray}
\sum_i{q_i} = Z
\\
\sum_i{m_i} = M_t
\\
\sum_i{s_i} = S
\\
\sum_i{x_i} = 1= \frac{1}{K}\sum_i{k_i}
\\
\sum_i{2n_i+|m_i| +1} \le N_{max}
\end{eqnarray}
where, for example, $k_i$ is the integer that defines the PBC longitudinal modes of Eq. \ref{Eq:longitudinal1} for the $i^{th}$ parton. The range of the number of fermion-antifermion pairs and bosons is limited by the cutoffs in the modes ($K$ and $N_{max}$).  Since each parton carries at least one unit of longitudinal momentum,  the basis  is limited to $K$ partons.  Furthermore, since each parton carries at least one oscillator quanta for transverse motion, the basis is also limited to $N_{max}$ partons.  Thus the combined limit on the number of partons is  $\min({K},{N_{max}})$.  In principle, one may elect to further truncate the many-parton basis by limiting the number of fermion-antifermion pairs and/or the number of bosons but we have not elected to do so here.

We may  refer to the quantity $K$ as the inverse longitudinal harmonic resolution.  We reason that as we increase $K$, higher longitudinal momenta states become available to the partons, thus allowing finer detail in the features of the longitudinal coordinate structure to emerge.

In a fully interacting application, the actual choice of symmetry constraints will depend on those dictated by the Hamiltonian.  For example, with QCD we would add color and flavor attributes to the single particle states and apply additional symmetries such as requiring all many-parton states to be global color singlets as discussed below.  Another example, which we adopt for the interacting QED example below, is the choice to conserve total $M_j=M_t + S$ rather than conserving $M_t $ and $S$ separately.  It is straightforward, but sometimes computationally challenging, to modify the symmetries in a basis function approach such as we adopt here.  However, in order to approach the continuum limit (all cutoffs are removed) as closely as possible with limited computational resources, one works to implement as many of the known symmetries as possible.

\subsection{Basis space dimensions}

For our defined non-interacting cavity mode problem, we now illustrate the exponential  rise in basis-space dimensions with increasing $N_{max}$ at fixed $K$, with increasing $K$ at fixed $N_{max}$ and with simultaneous increase in both cutoffs.  The first two situations involve a parton number cutoff defined by $K$ and $N_{max}$ respectively. Only the case with simultaneous increase in cutoffs keeps the problem physically interesting at higher excitations since this is the only case with unlimited number of partons as both cutoffs go to infinity.

\begin{figure}[tb]
\includegraphics[width=0.9\columnwidth]{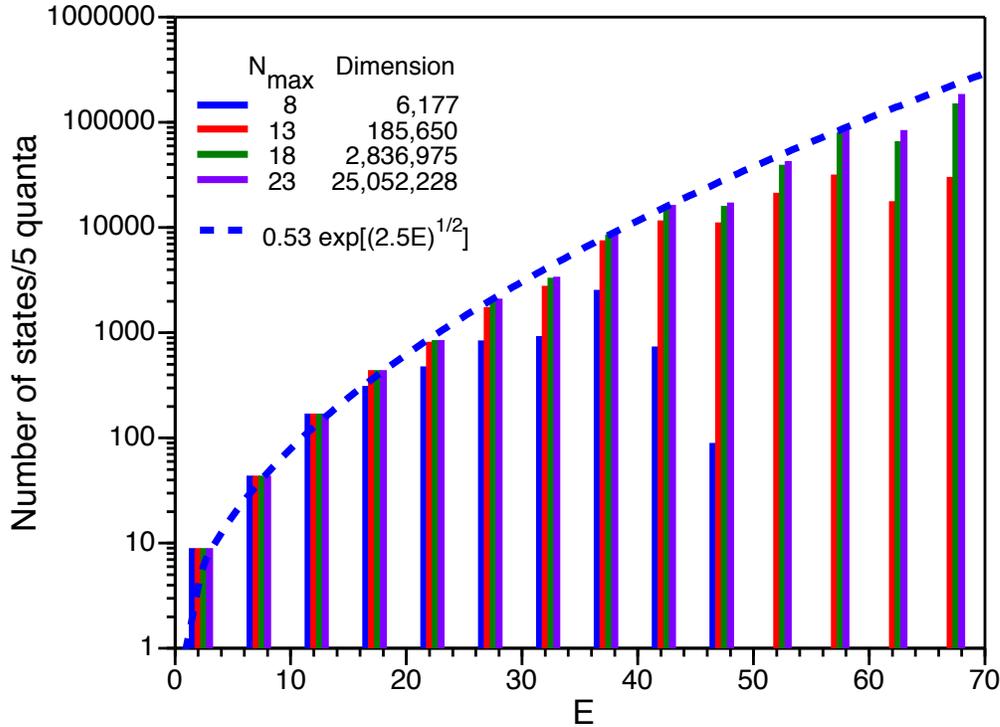}
\caption{(color online) State density as a function of dimensionless state energy E from BLFQ for non-interacting QED in a trap with no net charge and for a selection of $N_{max}$ values at fixed $K=6$.  The dimensions of the
resulting matrices are presented in the legend.  The states are binned in groups of 
5 units of energy (quanta) where each parton carries energy equal to its 2-D oscillator quanta 
($2n_i+|m_i|+1$) divided by its light-front momentum fraction ($x_i=k_i/K$). The dashed line traces an exponential in the square root of energy that reasonably approximates the histogram at larger $N_{max}$ values.} 
\label{Beq0_histogram}
\end{figure}

\begin{figure}[tb]
\includegraphics[width=0.9\columnwidth]{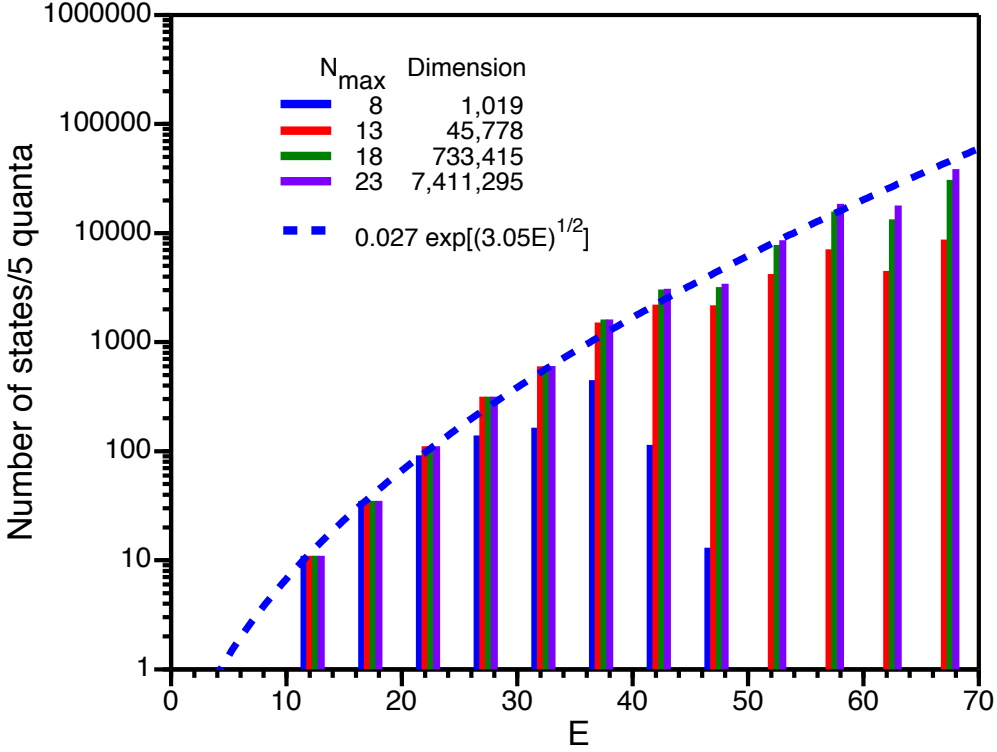}
\caption{(color online) State density as a function of dimensionless state energy E from BLFQ for non-interacting QED in a trap with net charge of 3 and for a selection of $N_{max}$ values at fixed $K=6$ .  The dimensions of the
resulting matrices are presented in the legend.  The states are binned in groups of 
5 units of energy (quanta) where each parton carries energy equal to its 2-D oscillator quanta 
($2n_i+|m_i|+1$) divided by its light-front momentum fraction ($x_i=k_i/K$). The dashed line traces an exponential in the square root of energy that reasonably approximates the histogram at larger $N_{max}$ values.} 
\label{Beq1_histogram}
\end{figure}

In Fig. \ref{Beq0_histogram} we present the state density in BLFQ for massless QED in the zero coupling limit for the case with no net charge $Z=0$, i.e. for zero lepton number.  Thus the cavity is populated by many-parton states consisting of fermion-antifermion pairs and photons.   The chosen symmetries are $M=0$ and $S=0$.  We show results for  $K=6$ at various values of $N_{max}$ spanning a range ($N_{max}=8-23$) .   The states are grouped to form a histogram according to their energy calculated from the chosen Hamiltonian in Eq. \ref{Hamiltonian} where we omit the constant preceding the summation for simplicity. Similarly, in Fig. \ref{Beq1_histogram} we present the state densities for $Z=3$, $M_t=0$ and $S=1/2$ at the same selected values of $N_{max}$.

Both Figs. \ref{Beq0_histogram} and \ref{Beq1_histogram} demonstrate the saturation of low-lying modes with increasing $N_{max}$.  That is, in each case, one may observe an excitation energy at which the state density reaches a value that no longer changes with increasing $N_{max}$.  The energy at which this saturation occurs, increases with $N_{max}$. We show only the lower sections of some of the state density distributions but it is clear that all distributions must fall off at sufficiently high energy for fixed $N_{max}$ and $K$.

\begin{figure}[tb]
\includegraphics[width=0.9\columnwidth]{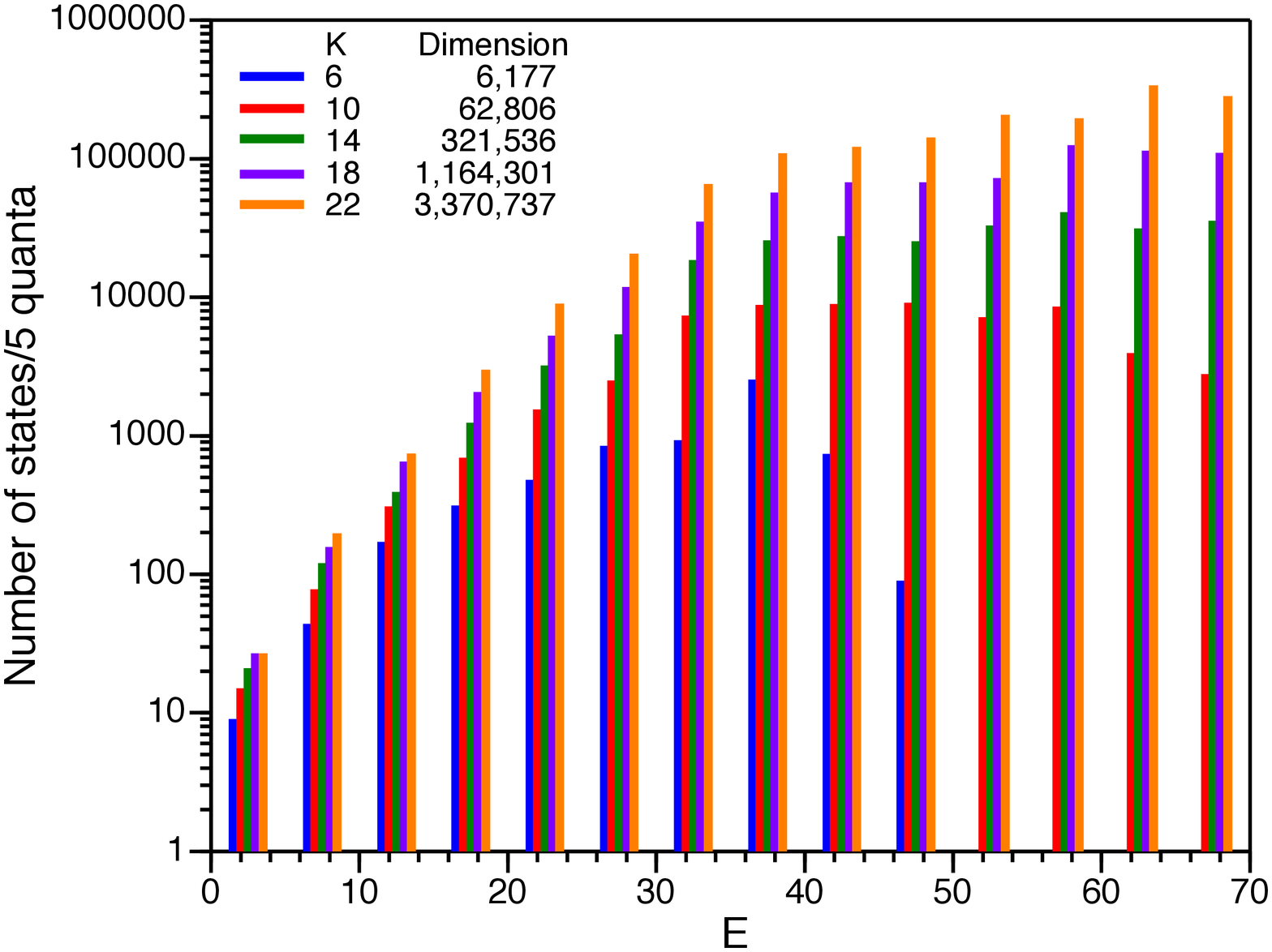}
\caption{(color online) State density as a function of dimensionless state energy E from BLFQ for non-interacting QED in a trap with no net charge and for a selection of $K$ values at fixed $N_{max}=8$.  The dimensions of the
resulting matrices are presented in the legend.  The states are binned in groups of 
5 units of energy (quanta) where each parton carries energy equal to its 2-D oscillator quanta 
($2n_i+|m_i|+1$) divided by its light-front momentum fraction ($x_i=k_i/K$). 
}
\label{Beq0_histogram_Kdep}
\end{figure}

In Fig. \ref{Beq0_histogram_Kdep} we present the state density in BLFQ for QED in the zero coupling limit again for the case with no net charge $Z=0$ but with increasing $K$ at fixed $N_{max}=8$.  In this case the many-parton states at low energy continue to increase in number with increasing 
$K$.  This is understandable from the definition of the Hamiltonian in Eq. \ref{Hamiltonian}.  In particular, a typical fermion-antifermion state with each parton's light-front momentum fraction close to $x_i = \frac{1}{2}$ achieves a low energy.  Correspondingly, as one increases $K$, the population of states at low E grows since there are more pairs of values of $x_i$ near $\frac{1}{2}$ to employ for minimizing the energy.  This reasoning easily extends to states with increasing numbers of partons so the net result is an increasing level density with increasing $K$ at fixed low E and fixed $N_{max}$.

\begin{figure}[tb]
\includegraphics[width=0.9\columnwidth]{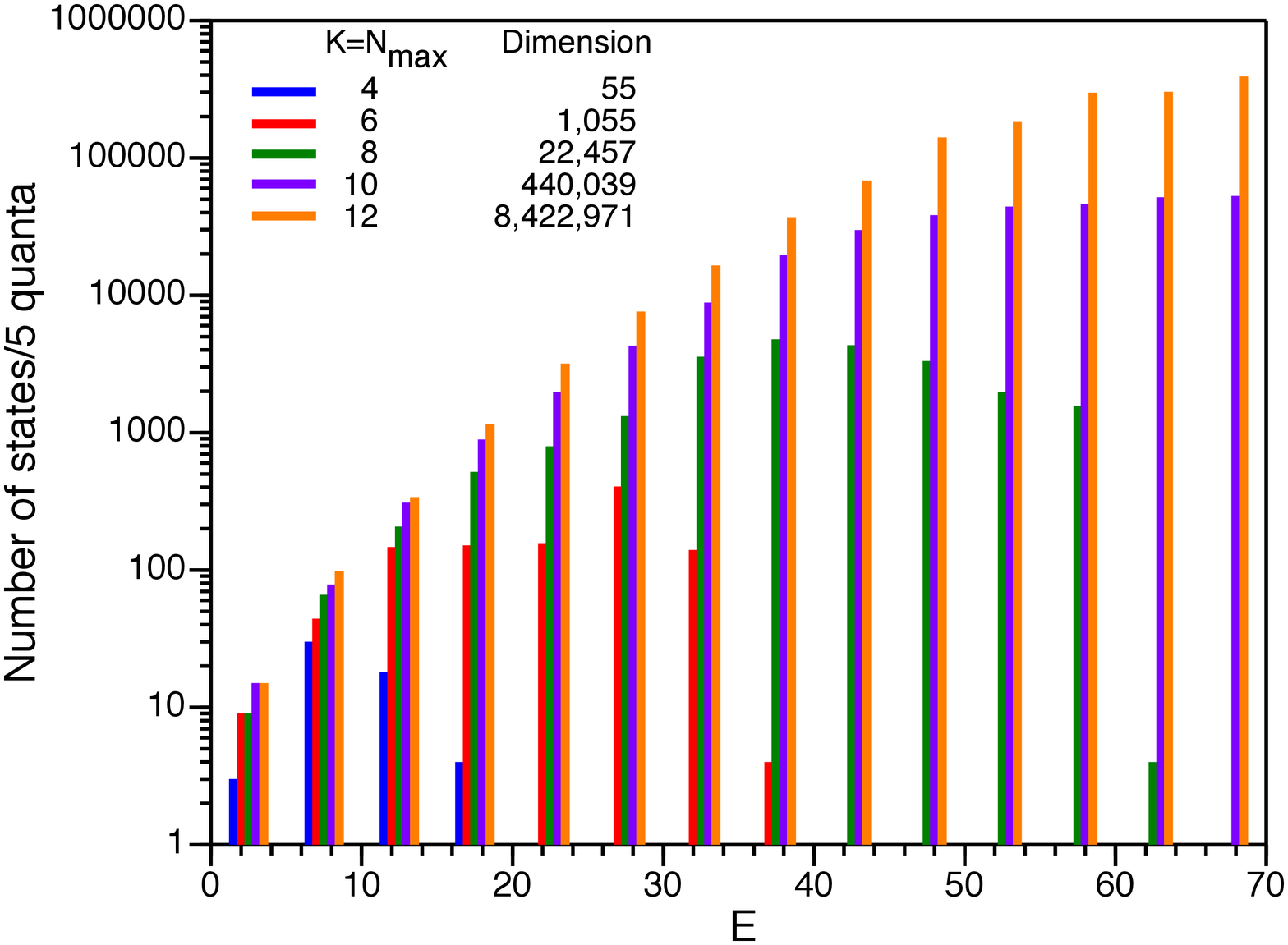}
\caption{(color online) State density as a function of dimensionless state energy E from BLFQ for non-interacting QED in a trap with no net charge and for $K=N_{max}$.  The dimensions of the resulting matrices are presented in the legend.  
The states are binned in groups of 5 units of energy (quanta) 
where each parton carries energy equal to its 2-D oscillator quanta 
($2n_i+|m_i|+1$) divided by its light-front momentum fraction ($x_i=k_i/K$). 
}
\label{Beq0_histogram_KNdep}
\end{figure}

For the final example of state densities, we consider the case where both $K$ and $N_{max}$ increase simultaneously.  For simplicity, we remain with the $Z=0$ sector and take $K = N_{max}$.  The state densities for this example are presented in Fig. \ref{Beq0_histogram_KNdep}.  Here, we observe trends similar to those shown in Fig. \ref{Beq0_histogram_Kdep} where there is no saturation in state density at low energy.

We take three cases depicted in Fig. \ref{Beq0_histogram_KNdep} to illustrate the distribution of many-parton states over the sectors of the Fock-space.  The distributions for the $N_{max}=K=8, 10$ and $12$ examples are  shown in Table \ref{tab_Beq0_KN8}.  With increasing cutoff, there is a rapid growth in the number of basis states within each Fock space sector.  Overall, there is approximately a factor of 20 increase in the total many-parton basis states with each increase of 2 units in the coordinated cutoff.

\begin{table}[ht]
\renewcommand{\arraystretch}{1.3} 
\begin{tabular}{|c|c|c|c|c|c|c|c|c|c|c|c|c|}
\hline
 $f \bar f$ pairs / bosons & 0  & 1 &  2 & 3 &  4  &  5  &  6  &  7  &  8  & 9  &  10 & Total\\ 
\hline
0     &  0 &  0 &  210  & 0 &  ~1122~  &  0  &  ~67~        &  0  &        1  &  ~~0~~  & ~~0~~ &   1400  \\
       &  0 &  0 &  495  & 0 &   ~11318 &  0  &   ~2936~   &  0  &      69 &       0       &     1      & 14819 \\
       &  0 &  0 & 1001 & 0 &     73600 &  0  &    63315    &  0  &  4027 &       0      &     69    &  142013 \\
 \hline
1    &  ~420~ &   1932 &    8190  &      1040 &  ~588~  &  8         &  2         &  0       & 0      &  0  &  0  & 12180     \\
      &  ~990~ & 10512 &  86856  &    33632 &  36672  &  1604  & 640     &  8       & 2      & 0  &  0  & 170916  \\
      &   2002  & 40810 & 574860 & 503040 & 929064 & 99962 & 60518 & 1770 & 644 & 8  &  2  & 2212680 \\
\hline
2     &  ~5961~ &    1560 &     1133    &      4        &      1        &        0  &       0  &  0  & 0 &  0  &  0  &    8659       \\
       &  64240   &   59240 &  97584    &      4040 &       1513 &       4  &       1  &  0  & 0 &  0  &  0  &   226622    \\
       & 427730 & 942240 & 2806624 & 381608 &  249825 & 4928 &  1565 & 4  & 1 &  0  &  0  & 4814525    \\
\hline
3     & 218       &     0         &        0      &      0      &  ~~0~~  &  ~~0~~  &  ~~0~~  &  ~~0~~  & ~~0~~ &  0  &  0 &          218 \\
       & 25584   & 1528      &      554   &      0      &  ~~0~~  &  ~~0~~  &  ~~0~~  &  ~~0~~  & ~~0~~ &  0  &  0 &      27666 \\
       & 808034 & 222336 & 200676 &   2592  &    602     &  ~~0~~  &  ~~0~~  &  ~~0~~  & ~~0~~ &  0  &  0 & 1234240 \\
\hline

4     &  ~~0~~  & ~~0~~ & ~~0~~ & ~~0~~ &  ~~0~~ &  ~~0~~  &  ~~0~~  &  ~~0~~  & ~~0~~ &  0  &  0  &         0  \\
       &     16      & ~~0~~ & ~~0~~ & ~~0~~ &  ~~0~~ &  ~~0~~  &  ~~0~~  &  ~~0~~  & ~~0~~ &  0  &  0  &        16 \\
       &  19325  &   168    &   20      & ~~0~~ &  ~~0~~ &  ~~0~~  &  ~~0~~  &  ~~0~~  & ~~0~~ &  0  &  0  & 19513 \\
\hline

\end{tabular}
\caption{Number of many-parton basis states in each Fock-space sector 
for two of the $N_{max}=K$ cases depicted in Fig. \ref{Beq0_histogram_KNdep}.
The counts are organized according to the number of fermion-antifermion 
($f \bar f$) pairs and the number of bosons in each sector. 
The first line in each $f \bar f$ row corresponds to the $N_{max}=K = 8 $ case which has
a total of 22,457 states, while the second line corresponds to the $N_{max}=K = 10 $ case
which has a total of 440,039 states. The third  line in each $f \bar f$ row corresponds to the $N_{max}=K = 12$ case which has a total of 8,422,971 states.  In this last case, there is a single 12-boson state not listed to save space.  The last column provides the total for that row.}
\label{tab_Beq0_KN8}
\end{table}

Specific cases in Table \ref{tab_Beq0_KN8} where no basis states may appear in a given Fock space sector may seem puzzling at first.  However, they are understandable once the symmetries and constraints are examined.  For example, with $N_{max}=K=8$ there are no states with 4 $f \bar f$ pairs since the Pauli principle excludes more than 2 pairs from occupying the lowest $N_{max}$ and $K$ modes.  Since two $f \bar f$ pairs must be in higher modes, either the total $K = 8$ or $N_{max}=8$ constraint will be violated by having a total of  4 $f \bar f$ pairs.

All our level density results are shown as a function of the dimensionless energy.  For the non-interacting theory in BLFQ only the kinetic term of the Hamiltonian contributes and the scale is available through an overall factor $\Lambda^2 = 2M_0\Omega$ as described above.  Without interactions and the associated renormalization program, one cannot relate the scales at one set of $(K,N_{max})$ values to another.  Ultimately, one expects saturation will arise with interaction/renormalization physics included as one increases the set of $(K,N_{max})$ values.

These state densities could serve as input to model the statistical mechanics of the system treated in the microcanonical ensemble.  Of course, interactions must be added to make the model realistic at low temperatures where correlations are important.  After turning on the interactions, the challenge will be to evaluate observables and demonstrate convergence with respect to the cutoffs ($N_{max}$ and $K$).  Independence of the basis scale, $\Omega$, must also be obtained.  These are the standard challenges of taking the continuum limit.  We will address these topics in a separate investigation.  For the current effort, we present a smooth representation for selected histograms, an exponential fit adopted from the well-known Bethe formula,  

\begin{eqnarray}
\rho (E) = b \exp(\sqrt{a E}),
\end{eqnarray}
where the precise values of the fitted constants are provided in the legends.  We provide these exponential fits in Figs. \ref{Beq0_histogram} and \ref{Beq1_histogram} where the low-lying state density exhibits saturation.

\subsection{Distribution functions}

In order to illustrate the potential value of the BLFQ approach, we present light-front momentum distribution functions for two simple toy models, based on results presented in Fig.  \ref{Beq0_histogram_KNdep}.  In the first example, we consider a model for a weak coupling regime and, in the second example, we consider a model for strong coupling behavior.  In both cases we introduce a simple state that is an equally-weighted superposition of basis states.  In the weak coupling case, we retain all basis states below a cutoff ($E_{cut} = 25$) in the dimensionless energy scale of Fig. \ref{Beq0_histogram_KNdep} for a given value of $K = N_{max}$.   That is, we imagine a situation where only the low-lying unperturbed many-parton basis states mix equally to describe a low-lying physical state of a weakly-coupled physical system. In the strong coupling case we retain all basis states of Fig. \ref{Beq0_histogram_KNdep} with equal weights for a given value of $K = N_{max}$.  Here, we imagine the coupling is so strong as to overwhelm the unperturbed spectrum and to produce a simple low-lying physical state with equal admixtures of all available basis states.  These states, labeled $| \Psi_w \rangle$ and $| \Psi_s \rangle$, where the $w$ ($``s"$) represents ``weak" (``strong") respectively,  are written as normalized sums over their respective sets of many-parton basis states $| \Phi_j \rangle$ as

\begin{eqnarray}
| \Psi_a \rangle = \frac{1}{\sqrt{D_a}} \sum_j{  | \Phi_j  \rangle}.
\end{eqnarray}
where $``a"$ represents $``w"$ or $``s"$ and the sum runs over the $D_a$ respective many-parton states. For our present application to probability distribution functions, the phases of the individual terms in expansion are not relevant so we choose all of them to be positive for simplicity.

\begin{figure}[tb]
\centering
\includegraphics[width=0.7\columnwidth]{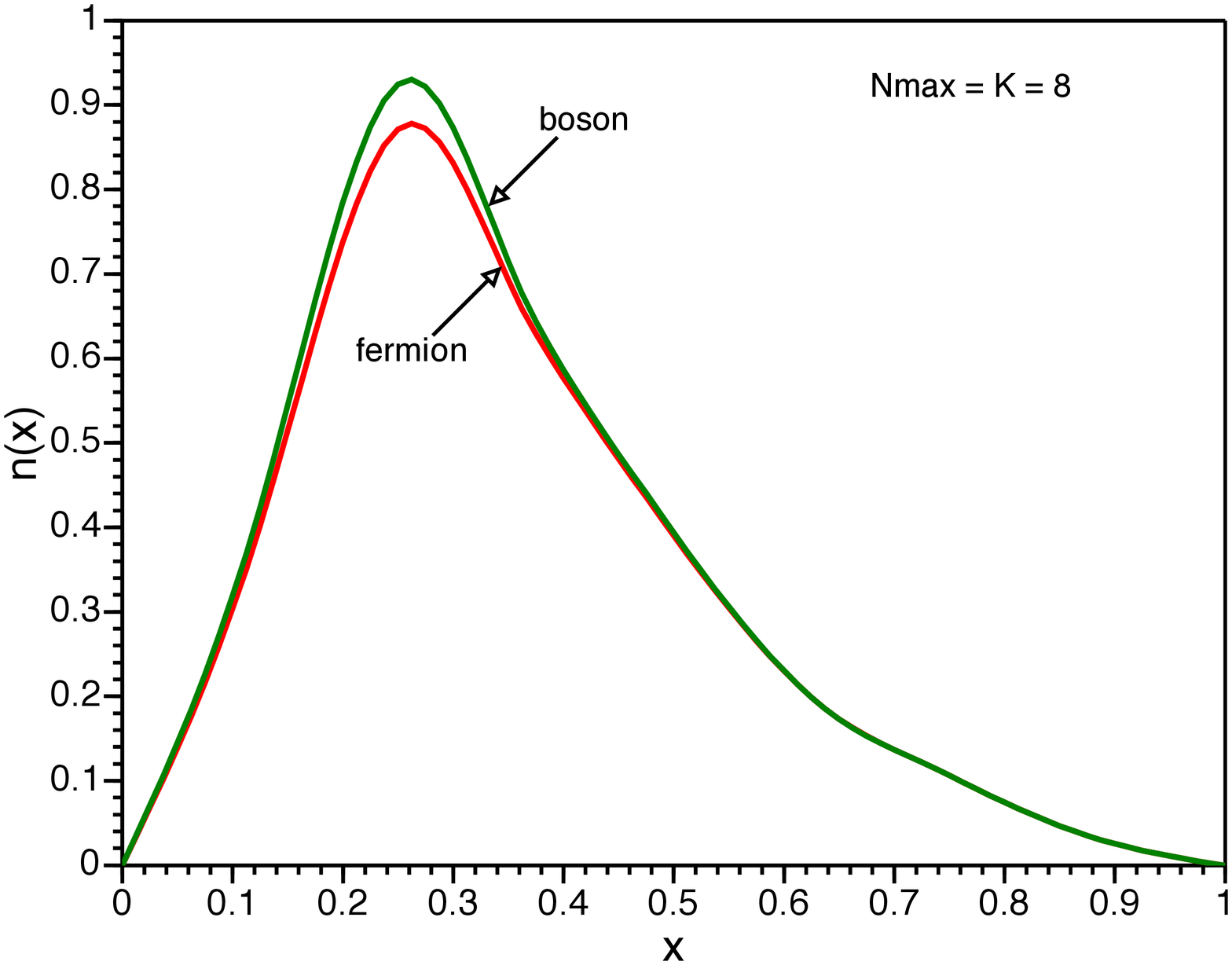}
\includegraphics[width=0.7\columnwidth]{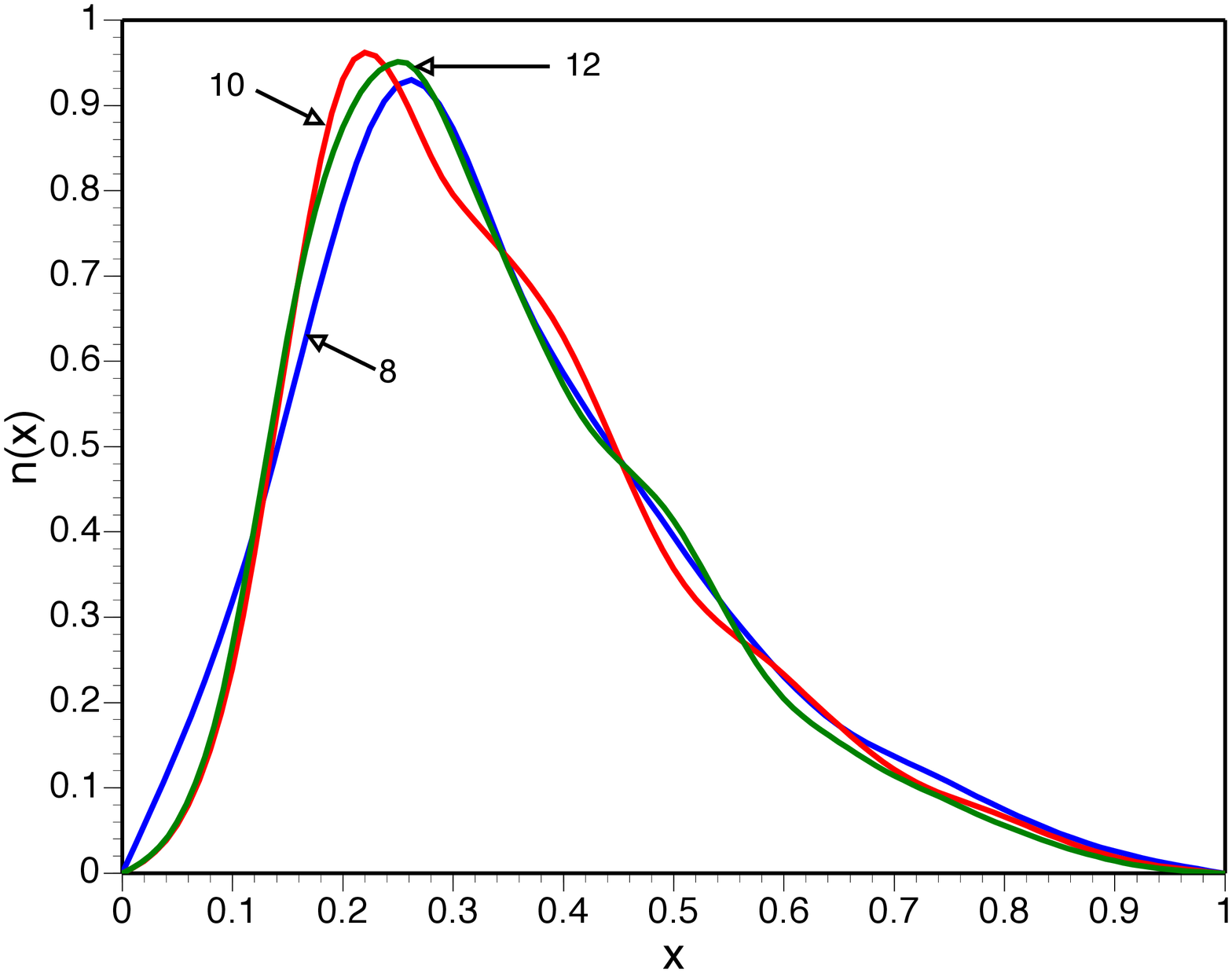}
\caption{(color online) Light front momentum distribution functions for states representing a weak coupling paradigm. The top panel displays the distributions at $N_{max}=K=8$.  The antifermion distribution is the same as the 
fermion distribution. The total momentum fraction carried by the fermion plus antifermion distribution is 0.66 while the boson distribution carries the remaining fraction 0.34.  The bottom panel displays the boson distributions at three different values of $N_{max}=K$ that are labeled.} 
\label{B0_parton_distributions_weak25}
\end{figure}

\begin{figure}[tb]
\centering
\includegraphics[width=0.7\columnwidth]{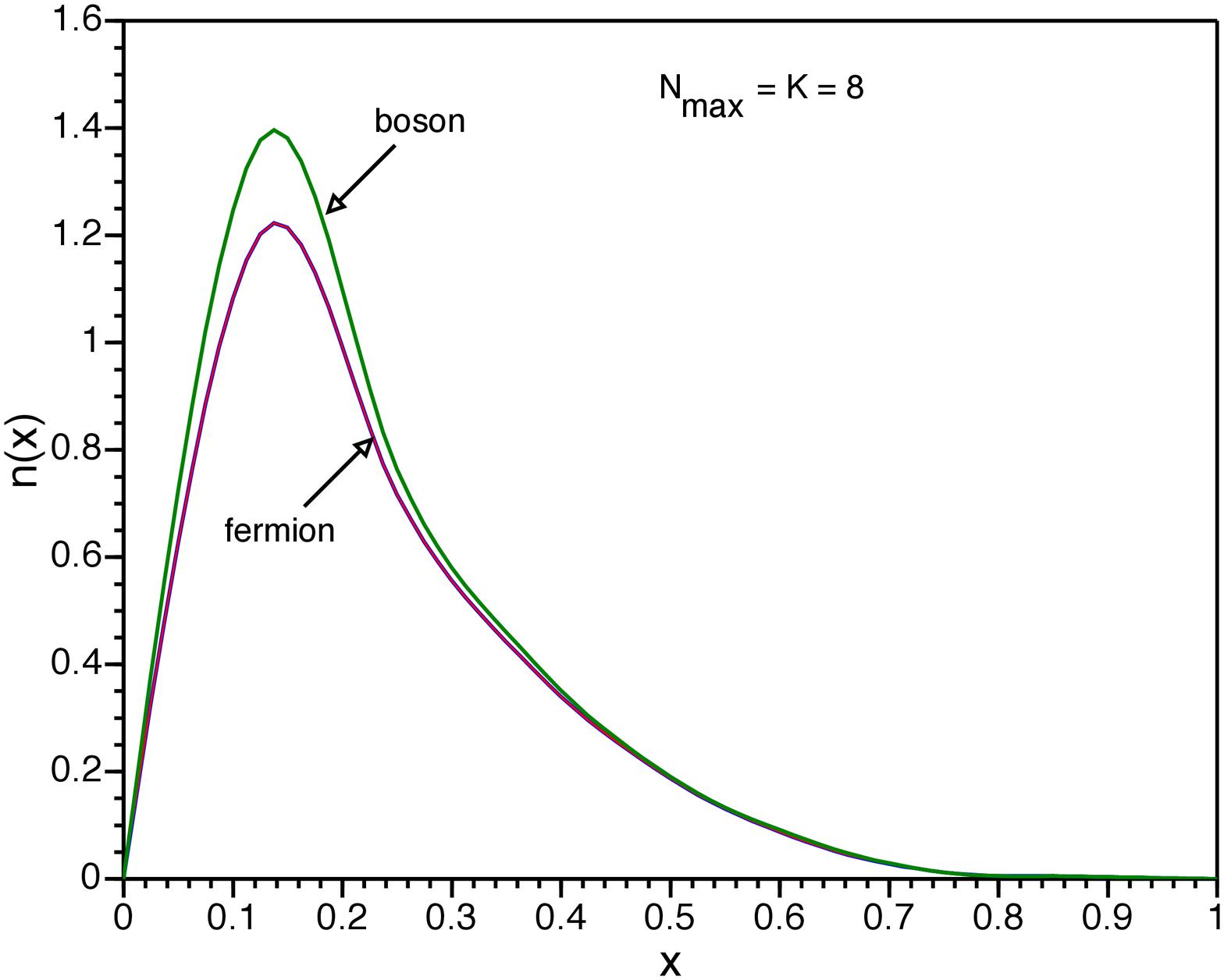}
\includegraphics[width=0.7\columnwidth]{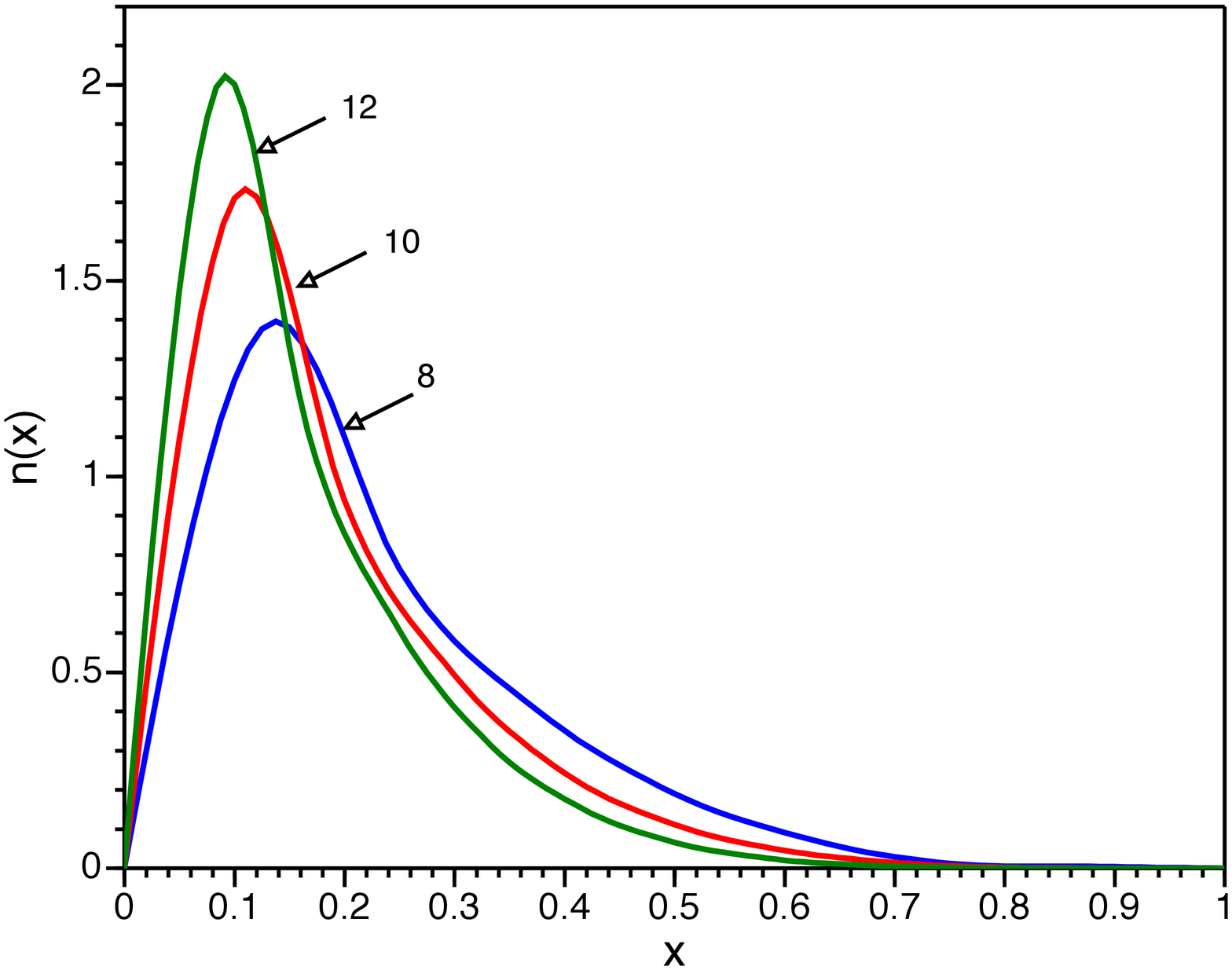}
\caption{(color online) Light front momentum distribution functions for states representing a strong coupling paradigm. The top panel displays the distributions at $N_{max}=K=8$.  The antifermion distribution is the same as the 
fermion distribution. The total momentum fraction carried by the fermion plus antifermion distribution is 0.65 while the boson distribution carries the remaining fraction 0.35.  The bottom panel displays the boson distributions at three different values of $N_{max}=K$ that are labeled.} 
\label{B0_parton_distributions_strong}
\end{figure}

Selected light-front momentum distributions $n(x)$ for these two model states are shown in Figs. \ref{B0_parton_distributions_weak25} and \ref{B0_parton_distributions_strong}.  The fermion and antifermion distributions are the same in these limiting examples.  Light-front momentum distributions are probability distributions emerging after integration over transverse degrees of freedom.  With our present selection of basis states, the light-front momenta take discrete values leading to discrete-valued distributions (histograms).  However, for convenience, we present smooth distributions in Figs. \ref{B0_parton_distributions_weak25} and \ref{B0_parton_distributions_strong} generated by spline interpolations.

The parton distributions at fixed $N_{max}=K$ satisfy both the normalization condition:
\begin{eqnarray}
\sum_i{  \int_0^1 {n_i (x) dx} } = \frac{1}{K}\sum_{i,k}{n_i(x_k)} = 1
\end{eqnarray}
and total light-front momentum conservation
\begin{eqnarray}
\sum_i{  \int_0^1 {x n_i (x) dx} } = \frac{1}{K}\sum_{i,k}{x_k n_i(x_k)} =1.
\end{eqnarray}
The index $i$ runs over the parton species (fermion, antifermion, boson) and the index $k$ runs over the discrete values of light-front momenta corresponding to the integers in Eq. \ref{Eq:longitudinal1} where $x_k = \frac{k}{K}$.  

The top panels of Figs. \ref{B0_parton_distributions_weak25} and \ref{B0_parton_distributions_strong} display the light-front momentum distributions at $N_{max}=K=8$ for the ``weak" and ``strong" coupling models, respectively.  The lower panels present the boson distribution functions for three $N_{max}=K$ values ranging from 8 to 12 for the same models.  

The fermion distributions are found to track the boson distributions with increasing $N_{max}=K$ and are not shown in the lower panels.  We also comment that the total momentum distribution fractions carried by the separate parton species appear approximately independent of $N_{max}=K$ over the range $8-12$.  About two-thirds of the total light-front momentum is carried by the fermions plus antifermions.  This division is characteristic of both the weak and strong coupling models over the $N_{max}=K = 8-12$ range we examined.

The top panel of Fig. \ref{B0_parton_distributions_strong} indicates a peak in the vicinity of the minimum light-front momentum fraction carried by a single parton in this basis, $x=\frac{1}{8}$, for both the fermions and the bosons.  This appears to be a characteristic of this strong coupling toy model and is illustrated in the lower panel of the same figure where the peaks in the boson light-front momentum distributions appear to track well with the inverse of $N_{max}=K$.  Clearly, with this toy model the distribution functions do not converge with increasing $N_{max}$ and $K$.

For comparison, we note that with the weak coupling toy model the peaks of the boson distributions shown in Fig. \ref{B0_parton_distributions_weak25} appear to be stable with increasing 
$N_{max}$ and $K$, and the distribution function appears to be reasonably well converged at 
$N_{max}=K=12$.  Based on these observations we anticipate good convergence for weakly interacting theories like QED.  

The lack of convergence of our strong coupling toy model may be worrisome for applications in QCD, but one should keep in mind that this toy model is far from realistic: all basis states are retained with equal weight.  Nevertheless, it is interesting to consider the trends of this model with increasing $N_{max}=K$.   For background, one may recall that deep-inelastic lepton scattering from a hadron in the scaling region $Q^2 \to \infty $ provides a measure of the hadron's charged quark distribution functions.  With more detailed resolution provided by the virtual photon exchange (increasing $Q$ leads to shorter wavelengths), experiments reveal that the charged quark distributions evolve to lower values of light-front momentum fraction, $x$.  The pattern shown in the lower panel of Fig. \ref{B0_parton_distributions_strong} with increasing $N_{max}=K$ is reminiscent of this experimental trend with increasing $Q$.  Given the simplicity of the strong interaction model, one may infer that the evolution of multi-parton phase space with increasing $N_{max}=K$ could play a significant role in the evolution of light-front momentum distribution functions with improved resolution through increasing $Q$.

\subsection{Extension to color}

We can extend the approach to QCD by implementing the SU(3) color degree of freedom for each parton - 3 colors for each fermion and 8 for each boson.  For simplicity, we restrict the present discussion to the situation where identical fermions occupy distinct space-spin single-particle modes.  The case where we allow multiple space-spin occupancies by identical fermions leads to color space restrictions.  We will address this additional complexity in a subsequent investigation.

We consider two versions of implementing the global color-singlet constraint for the restricted situation under discussion here.   In both cases we enumerate the color space states to integrate with each space-spin state of the corresponding partonic character. 

\begin{figure}[tb]
\centering
\includegraphics[width=0.45\columnwidth]{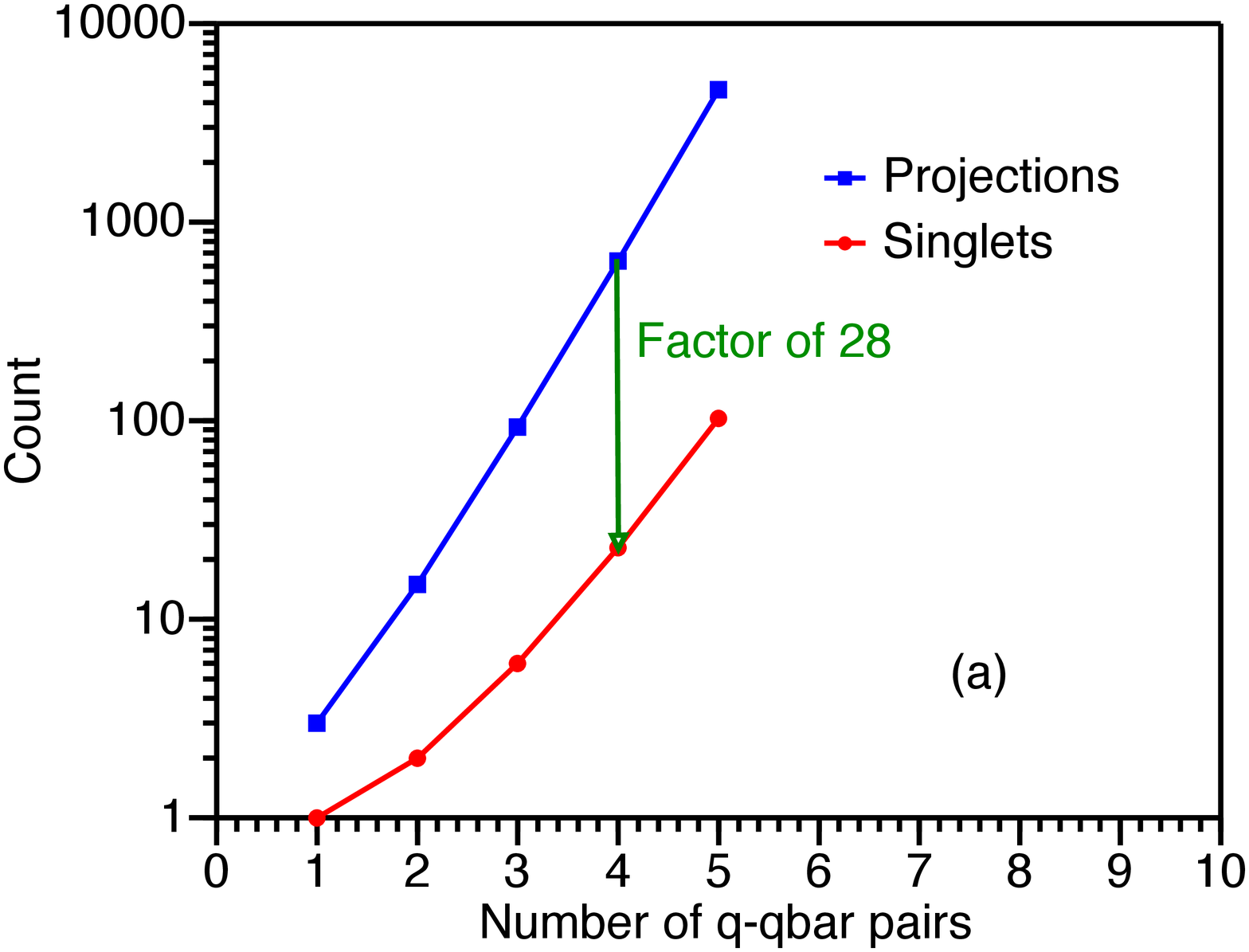}\includegraphics[width=0.45\columnwidth]{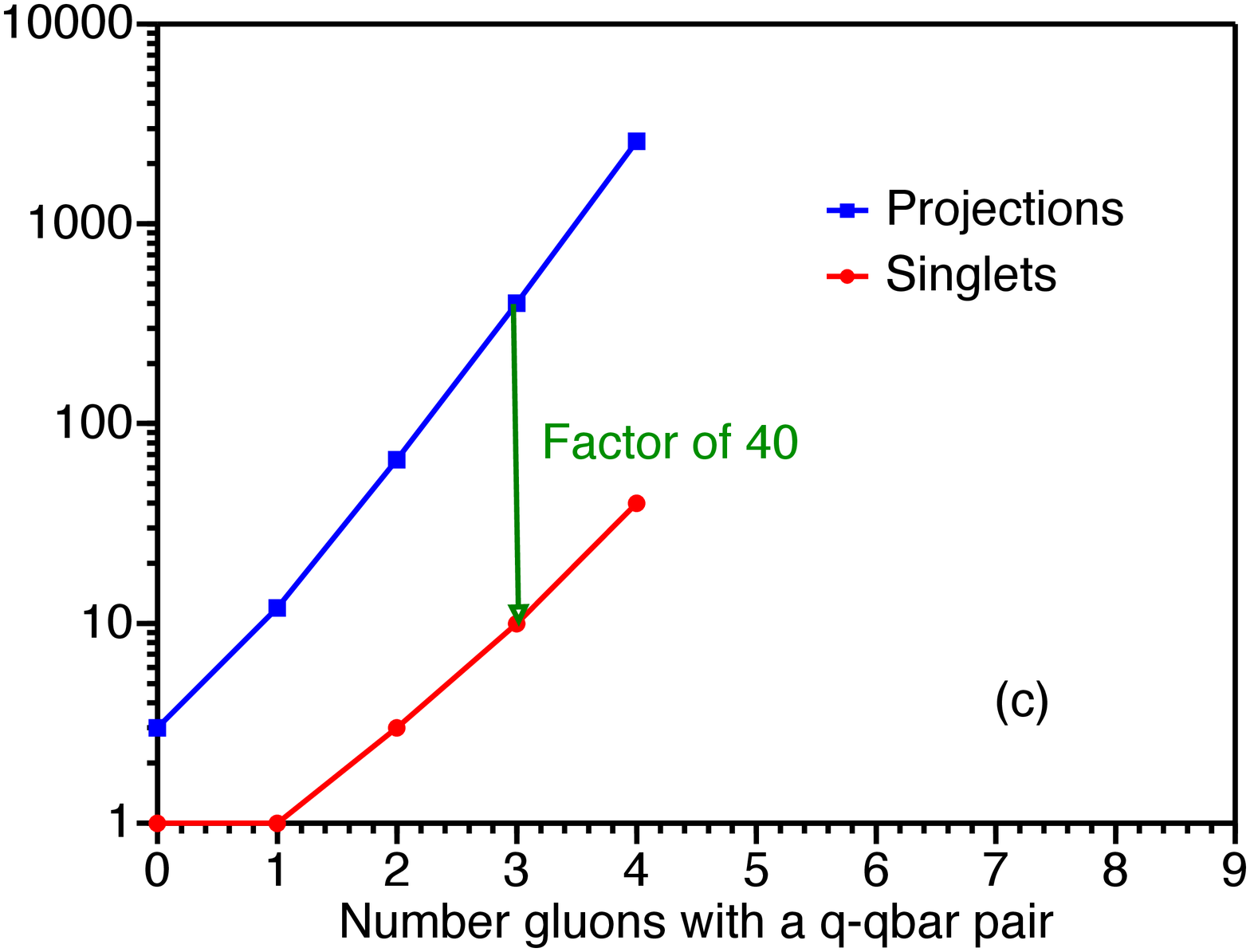}
\includegraphics[width=0.45\columnwidth]{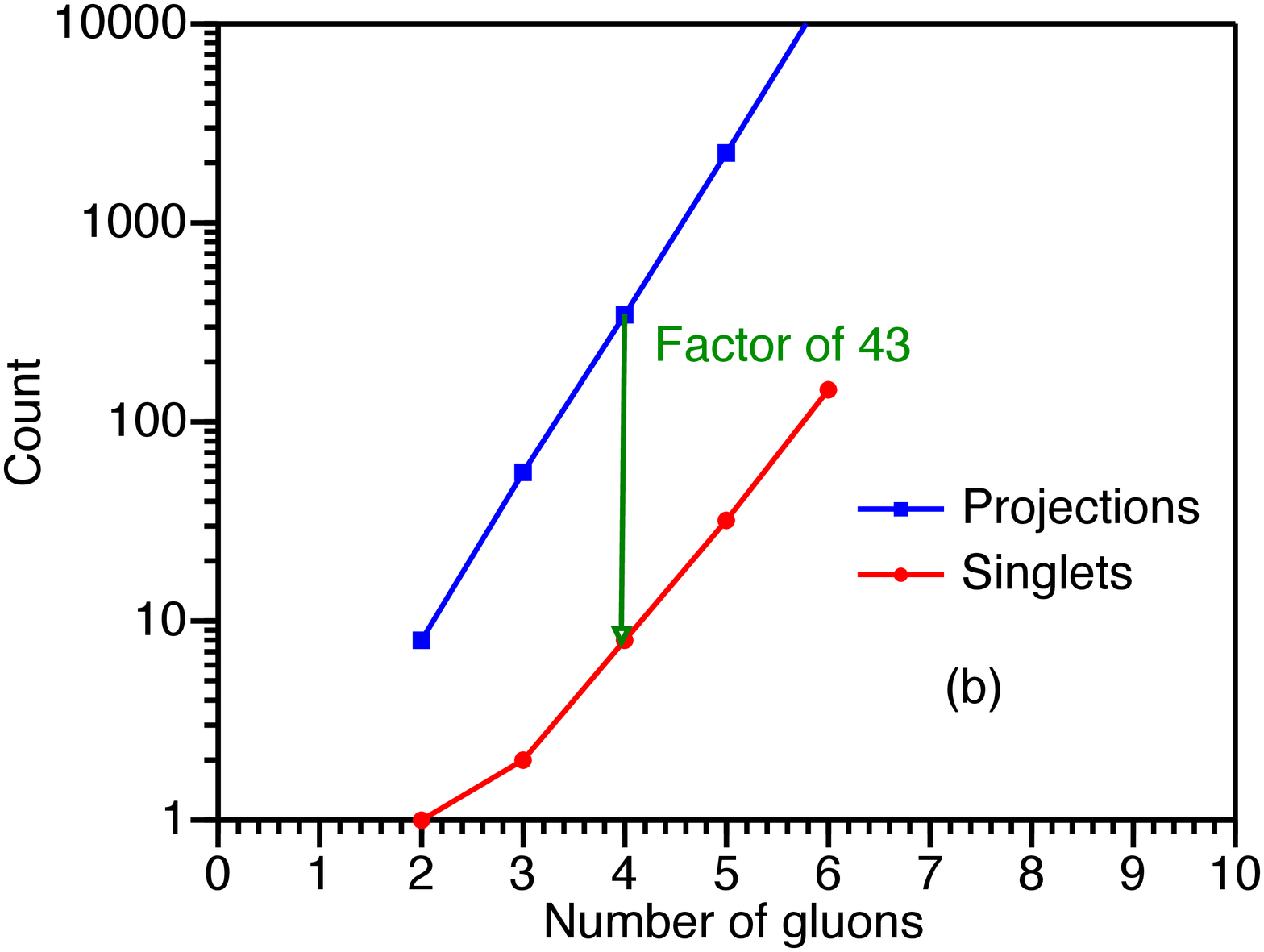}\includegraphics[width=0.45\columnwidth]{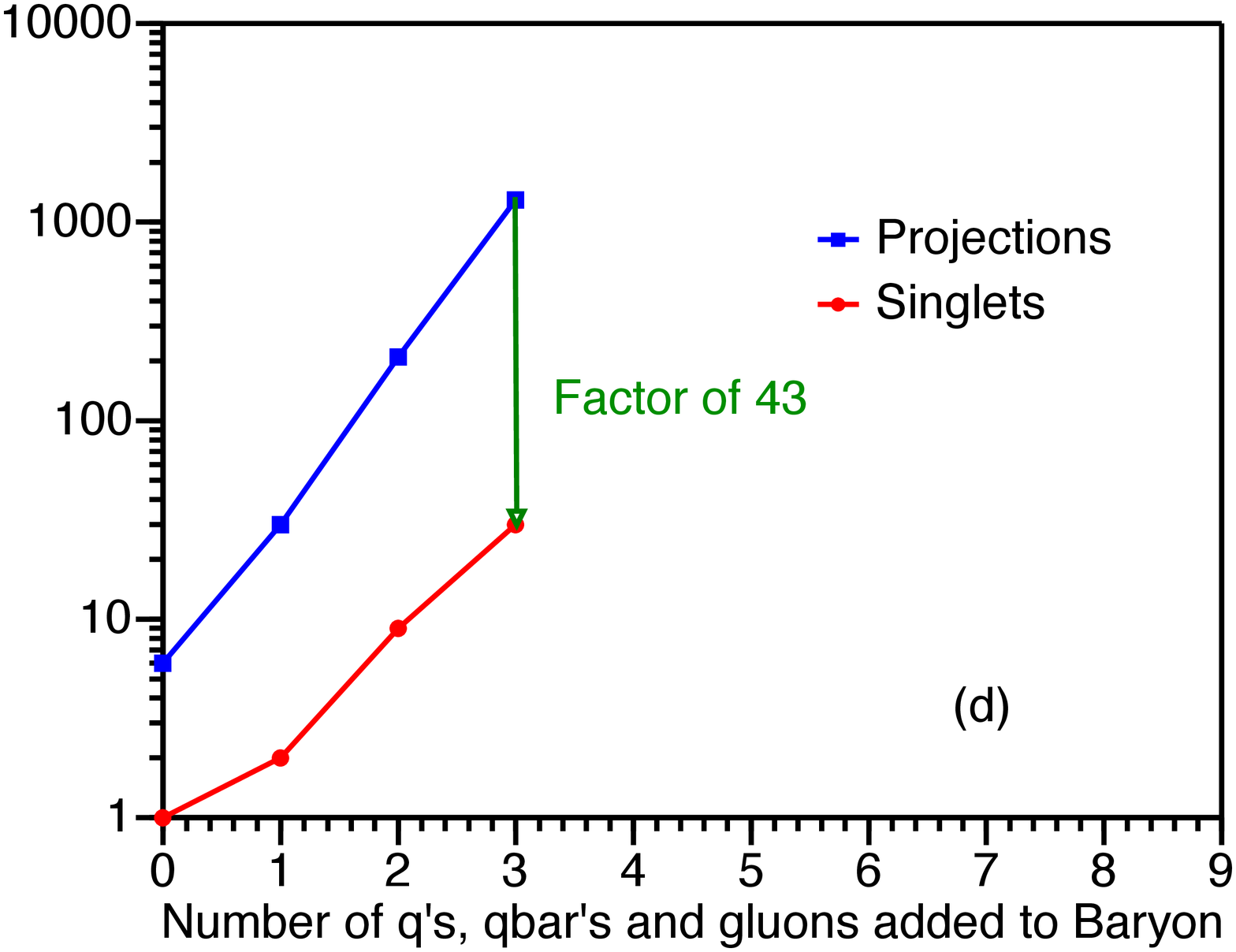}
\caption{(color online) Number of color space states that apply to each space-spin configuration of selected multi-parton states for two methods of enumerating the color basis states. The upper curves are counts of all color configurations with zero color projection.  The lower curves are counts of global color singlets.} 
\label{colorstates}
\end{figure}

In the first case, we follow Ref. \cite{Lloyd} by enumerating parton states with all possible values of SU(3) color. Thus each space-spin fermion state goes over to three space-spin-color states.  Similarly, each space-spin boson state generates a multiplicity of eight states when SU(3) color is included. We then construct all many-parton states having zero color projection.  Within this basis one will have both global color singlet and color non-singlet states.  The global color-singlet states are then isolated by adding a Lagrange multiplier term in many-parton color space to the Hamiltonian so that the unphysical color non-singlet states are pushed higher in the spectrum away from the physical color single states.  To evaluate the increase in basis space dimension arising from this treatment of color, we enumerate the resulting color-singlet projected color space states and display the results as the upper curves in Fig. \ref{colorstates}.

In the second case, we restrict the basis space to global color singlets and this results in the lower curves in Fig. \ref{colorstates}.  The second method produces a typical factor of 30-40 lower multiplicity at the upper ends of these curves at the cost of increased computation time for matrix elements of the interacting Hamiltonian.  That is, each interacting matrix element in the global color-singlet basis is a transformation of a submatrix in the zero color projection basis. Either implementation dramatically increases the state density over the case of QED, but the use of a global color-singlet constraint is clearly more effective in minimizing the explosion in basis space states.

We note that, for the pure multi-fermion basis space sector, shown in the upper left panel of Fig. \ref{colorstates}, we could have produced the lower curve using methods introduced and applied successfully in 1 + 1 dimensional QCD \cite{Hornbostel:1988fb}.  That is, the number of global color singlets for a given fermion-only basis state, with other (non-color) quantum numbers specified, is independent of the number of spatial dimensions provided there is at least one.

\section{Cavity mode light-front field theory - elements of the interacting theory}

Now we briefly address the interacting theory where a primary concern will be to manage the divergent structure of the theory.  There are two possible locations for divergences in a Hamiltonian basis function approach: (1) the matrix elements themselves diverge, or (2) the eigenvalues diverge as one or more cutoffs are removed.

In our cavity field theory applications with interactions, we manage these divergences with the help of suitable counterterms, coupling constant and mass renormalizations,  and  boundary condition selections.  The development of counterterms is expected to be straightforward as seen, for example, in Ref. \cite{JPV214}.  The infrared divergences in light-front momentum arising in both the fermion to fermion-boson vertex as well as in the instantaneous fermion-boson interaction are expected to be well-managed by previously defined counterterms \cite{JPV214} suitably transcribed for the transverse basis functions we have adopted.  We anticipate this prescription will work since the longitudinal modes we adopt are similar to those used in Ref. \cite{JPV214}.  Finally, we expect the transverse ultraviolet divergences to be suitably-managed with our basis function selection. 

Since we are introducing a basis-function approach for the transverse degrees of freedom, we need to investigate convergence rates with increased cutoff of the transverse modes $N_{max}$.   Here, in a simple set of examples, we outline how we can search for additional sources of divergence with the help of a perturbation theory analysis.

Consider the second order energy shift, $\Delta E$, induced on a single parton in the transverse mode ($n,m$) by its coupling $V$ to partons in higher energy transverse modes ($n',m'$): 

\begin{eqnarray}
\Delta E_{n,m} = \sum_{n',m'} {\frac{|\langle n,m |V| n',m' \rangle|^2}{E_{n,m} - E_{n',m'}}} \le 0
\\
\Delta E_{n,m} \approx \int{\frac{|\langle n,m |V| n',m' \rangle|^2}{E_{n,m} - E_{n',m'}} \rho (\bar n')d \bar n'},
\end{eqnarray}
where we use the following notation and properties of the 2-D harmonic oscillator
\begin{eqnarray}
\bar n' = 2n' + |m'|
\\
E_{n',m'} = (\bar n' + 1)  \Omega
\\
\rho (\bar n') = \bar n' + 1
\end{eqnarray}
and we converted the sum to an integral taking the degeneracy into account with $\rho$.

Thus, according to perturbation theory, we expect a UV divergence if the matrix element falls off too slowly with increasing $\bar n'$.  In particular, if the matrix element falls approximately as
 $(\bar n')^{-\frac{1}{2}}$, then we expect a logarithmic divergence since the integrand will have a net $(\bar n')^{-1}$ dependence.   If the falloff is even slower then we encounter a more serious divergence.
 
Another possible source of a log divergence could arise within the selected sum over $m'$ in which case $\rho$, the level density factor in the integrand, is unity. Then, if the matrix elements for fixed $n,n'$ are approximately constant with increasing $m'$, we again find a log divergence in the sum over $m'$.

For a first investigation, we have examined the behavior of various sets of matrix elements for the fermion to  fermion-boson vertex.   For the purpose of this investigation, we adopt periodic (antiperiodic) boundary conditions for the longitudinal modes of the bosons (fermions) and we hold spin projections fixed for initial and final states.  We then adopt specific values for the longitudinal momentum fractions observing the conservation rule.   The trends we examine should not be sensitive to the specific values adopted. 

To date, we have found no matrix element trends with increasing transverse energy that would imply new divergences.  All such matrix elements sets we examined, within a perturbative analysis, fall faster than the inverse square root of the principal quantum number $n'$.  Also, we found no sets that remained constant with increasing $m'$ (and thus $m$) holding  $n,n'$ fixed. 

\begin{figure}[tb]
\includegraphics[width=0.9\columnwidth]{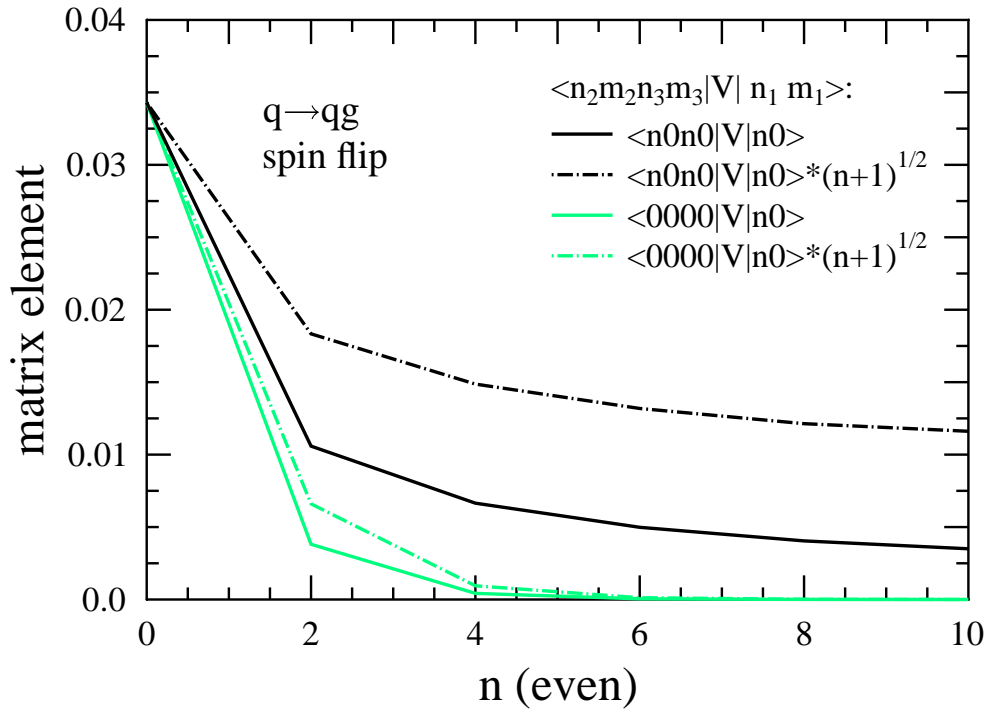}
\caption{(color online) Behavior of representative fermion to fermion-boson matrix elements in BLFQ. The quantum numbers specifying the parton transverse modes $(n_i,m_i)$ in the matrix elements are given in the legend. Only the transverse mode contributions to the matrix elements are shown.  Results are also shown with a multiplicative factor of $\sqrt{n+1}$ applied to help search for a logarithmic divergence by obtaining a resulting flat behavior, when it occurs. Overall matrix element normalization depends on the specific values of light-front momentum fractions carried by the interacting partons.} 
\label{n_even}
\end{figure}

\begin{figure}[tb]
\includegraphics[width=0.9\columnwidth]{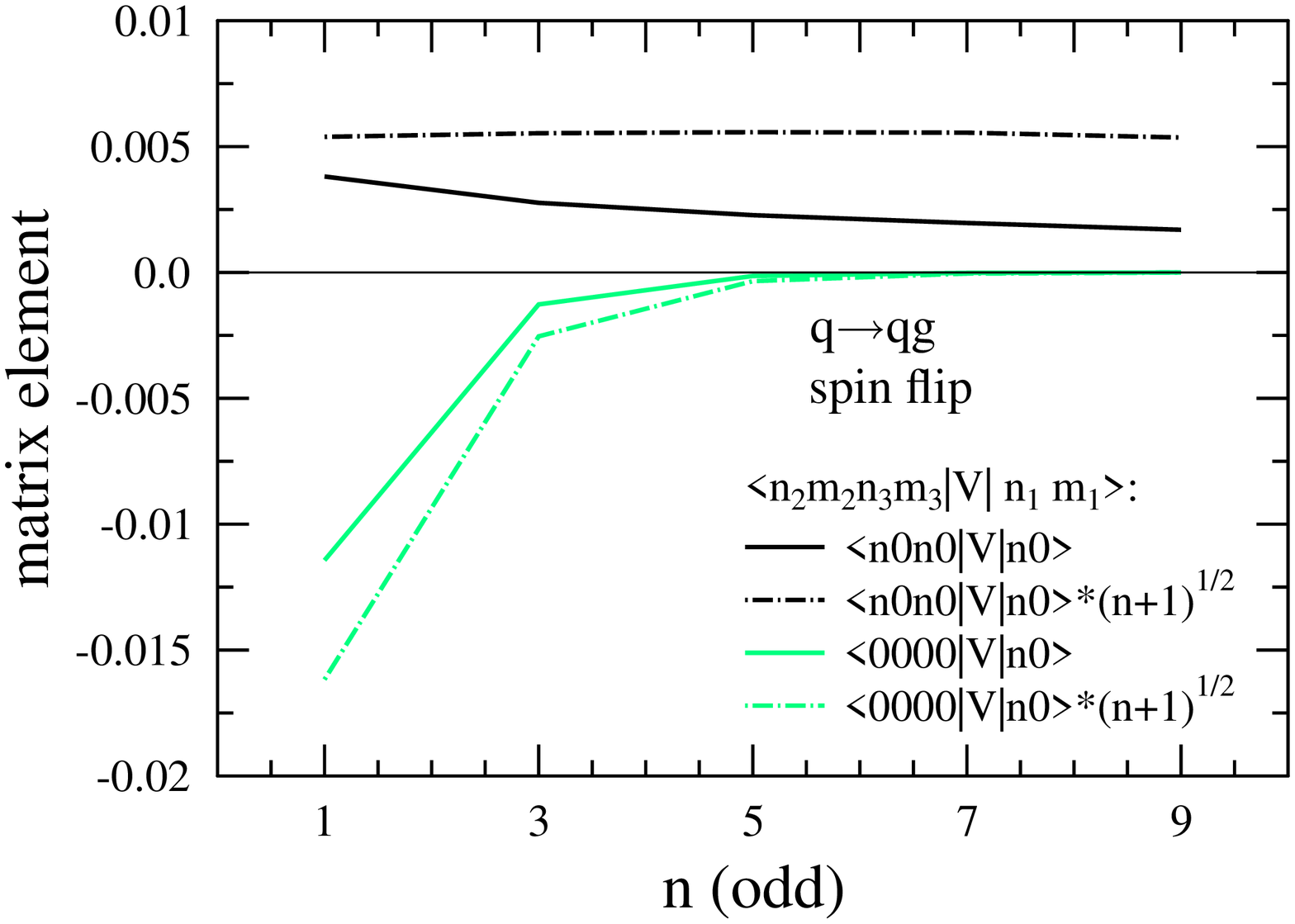}
\caption{(color online) Behavior of representative fermion to fermion-boson matrix elements in BLFQ. The quantum numbers specifying the parton transverse modes $(n_i,m_i)$ in the matrix elements  are given in the legend.  Only the transverse mode contributions to the matrix elements are shown.  Results are also shown with a multiplicative factor of $\sqrt{n+1}$ applied to identify logarithmic divergence by a resulting flat behavior, when it occurs.  Note that for one of the cases shown, the resulting matrix elements vanish with increasing $n$ while the other case shows a constant trend but does not enter a second-order perturbative analysis. Overall matrix element normalization depends on the specific values of light-front momentum fractions carried by the interacting partons.} 
\label{n_odd}
\end{figure}

We portray in Figs. \ref{n_even} and  \ref{n_odd} representative sequences of how these off-diagonal matrix elements behave as one increases the difference in the initial and final state principal quantum numbers.  We also portray two interesting cases where the fermion and fermion-boson principal quantum numbers track each other,  cases that do not enter a perturbative analysis.  Note that we have limited the illustrations to the transverse components of our matrix elements. We also select  cases where the fermion spin is flipped, cases that are proportional to the fermion mass. We set the fermion mass to unity so the results are expressed in units of the fermion mass.

We further limit our presentation to the case where all partons remain in the orbital projection quantum number zero state and the 2-parton (fermion-boson) states have each parton in the same transverse state.  For the non-perturbative illustrative cases, we display the matrix element trends where all partons remain in the same transverse mode, $(n,0)$. 

When the single fermion state has an even value of the principal quantum number $n$ as shown in Fig.  \ref{n_even},  the matrix elements appear to be well-behaved either when the 2-parton configuration is the lowest accessible case ($\langle 0000 |$) or when the each of the two partons resides in the state with the same principal quantum number as the single fermion state.  We demonstrate anticipated good convergence with increasing $n$ by showing that the matrix elements, when multiplied by $\sqrt{n+1}$, still fall with increasing $n$.

For the case when the single fermion state has an odd value of the principal quantum number $n$ as shown in Fig.  \ref{n_odd}, the situation is somewhat different.  For the matrix element set entering a perturbative analysis, the matrix elements fall to zero with increasing $n$ sufficiently fast that multiplying by $\sqrt{n+1}$ does not significantly distort the trend to zero.  However, the large $n$ behavior of the fermion-boson matrix element, with all partons at the same $n$, is seen to go approximately as $\sqrt{n+1}$.  This is best seen in Fig.  \ref{n_odd} where the matrix elements are multiplied by $\sqrt{n+1}$ and the result appears to be a nonzero constant at large $n$.  Since this  trend does not appear in a second order perturbation theory analysis, we must await the full Hamitonian diagonalization in sufficiently large basis spaces to  better understand its role in the convergence with increasing $N_{max}$.

As a result of this initial analysis, we anticipate that straightforward adoption of counterterm methods previously introduced  \cite{JPV214} will be sufficient for managing the identified divergences in BLFQ.  However, it seems advisable to have an alternative scheme for comparison.   Therefore, we plan to adopt a second approach that involves a recently proposed sector-dependent coupling constant renormalization scheme \cite{Karmanov}.  Another alternative which we may adopt uses the Pauli-Villars
regulator \cite{Brodsky:1998hs}.

\begin{figure}[tb]
\includegraphics[width=0.8\columnwidth]{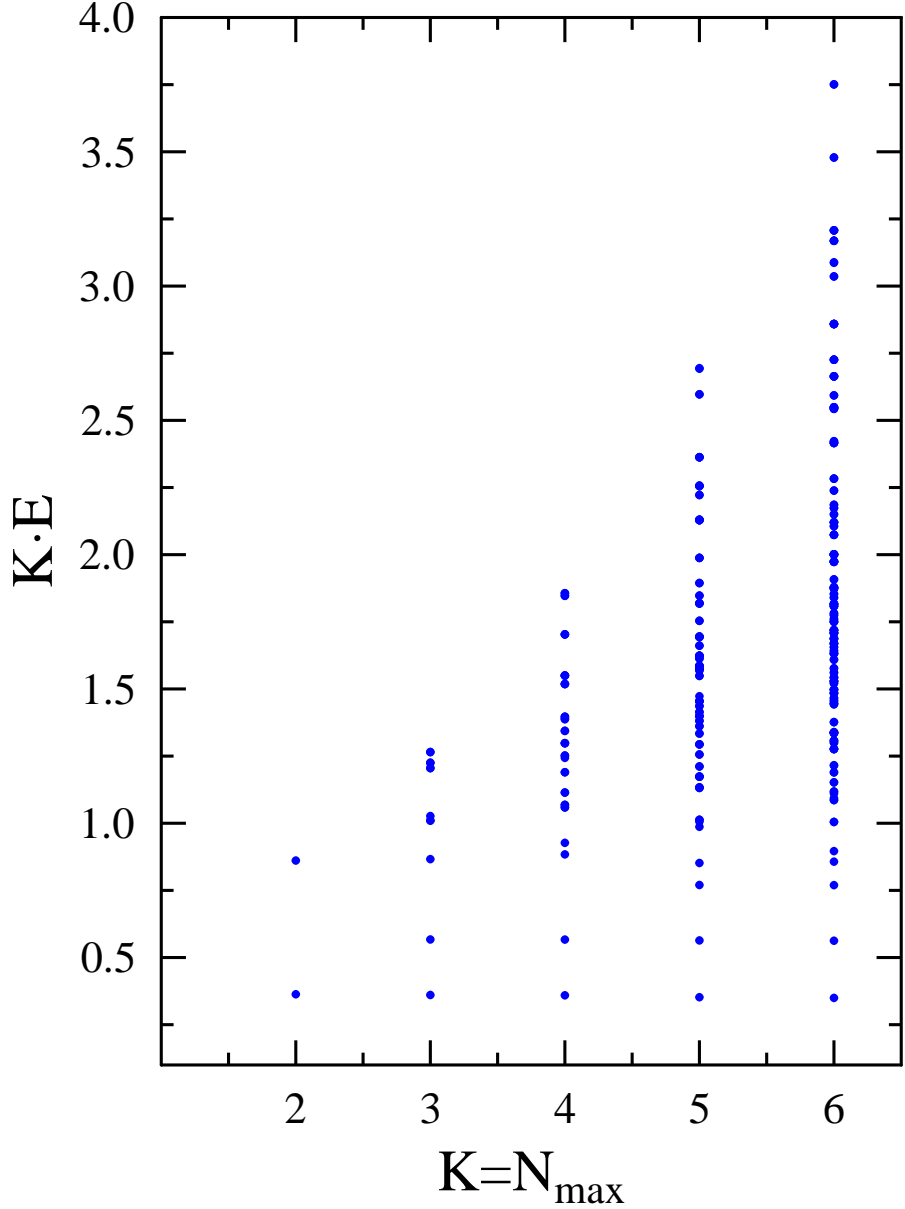}
\caption{(color online) Eigenvalues (multiplied by $K$) for a non-renormalized 
light-front QED Hamiltonian which includes the fermion-boson vertex and the 
instantaneous fermion-boson interaction without counterterms. The basis is limited to fermion and 
fermion-boson states satisfying the symmetries. The cutoffs
for the basis space dimensions are selected such that $K$ 
increases simultaneously with the $N_{max}$. The harmonic oscillator parameters were chosen as
$\Omega=0.1$ MeV and $M_0=0.511$ MeV, and the fermion mass was chosen to be 
equal to $M_0$. } 
\label{eigenv}
\end{figure}

Even without regulating the possible divergencies in the Hamiltonian, it is possible to get some
idea how  the cutoffs for the basis space dimensions, as discussed in the previous section, 
affect the eigenvalue spectra of the Hamiltonian. 
In Fig. \ref{eigenv} we show the eigenvalues (multiplied by $K$) for a light-front QED Hamiltonian 
in a basis space limited to the 
fermion and fermion-boson sectors.   For this particular example we chose
the harmonic oscillator parameters as $\Omega=0.1$ MeV and $M_0=0.511$ MeV, and the fermion mass was chosen to be equal to $M_0$. The interaction terms include the fermion to fermion-boson vertex and the instantaneous fermion-boson interaction. We chose the basis
space such that the basis states have total $M_j=M_t+S = \frac{1}{2}$, and we simultaneously
increase the $K$ and $N_{max}$ cutoffs.
As a result, the size of the Hamiltonian matrices increase rapidly. For $K=N_{max}=2,3,4,5,6$,
the dimensions of the corresponding matrices are $2\times 2,\, 12\times 12,\, 38\times 38,\,
 99\times 99$ and $208\times 208$ respectively. 
 
The number of the single fermion basis states increase slowly with increasing $K=N_{max}$ cutoff.  
For  $K=N_{max}=2,3,4,5,6$ the number of single fermion basis states is $1,2,2,3,3$, respectively.
Our lowest-lying eigenvalues correspond to solutions dominated by these states and they appear with nearly harmonic separations in  Fig. \ref{eigenv} as would be expected at the coupling of QED.

The higher eigenstates are the ones dominated by the fermion-boson basis states that interact
with each other in leading order through the instantaneous fermion-boson interaction. Their multiplicity
increases rapidly with increasing $K=N_{max}$ and they exhibit
significant mixing with each other as well as weak mixing with the lowest-lying states.   The eigenvalues dominated by the fermion-boson basis states cluster in nearly degenerate groups above the lowest-lying states.  
 
Further progress requires that we implement our renormalization program as well as a major expansion in the basis space size.  These efforts are underway and will be presented in a subsequent paper.

\section{Conclusions and Outlook}
Following successful methods of {\it ab initio} nuclear many-body theory, we have introduced a basis light-front quantization (BLFQ) approach to Hamiltonian quantum field theory and illustrated some of its key features with a cavity mode treatment of massless non-interacting QED.  

Cavity mode QED, with a 2D harmonic oscillator for the transverse modes and longitudinal modes chosen with periodic boundary conditions, exhibits the expected dramatic rise in many-parton basis states as the cutoffs are elevated. With the non-intracting cavity-mode Hamiltonian, we obtain the state density distributions at various choices of the regulators.  These basis state densities provide initial elements of a quantum statistical mechanics approach to systems treated in the BLFQ approach. We then illustrated the access to light-front momentum distribution functions in this approach with simple models of wavefunctions that reflect possible interaction effects.

In order to extend our method to QCD, we have evaluated two methods for treating the color degree of freedom. Since large sparse matrices will emerge, we argue that it is more efficient in storage requirements to adopt multi-parton basis states that are global color singlets and we presented sample measures of the efficiency gains over basis states with color-singlet projection alone.   To achieve this savings in storage (reduced matrix size) we will incur an increase in the computational effort for the non-vanishing matrix elements.

We have also outlined our approach to managing the expected divergences that will preserve all the symmetries of the theory. An initial inspection of the interaction vertices of QED in the BLFQ approach shows smooth behaviors that, following a second-order perturbative analysis, are not expected to lead to divergences.  It appears that the cavity mode treatment, with the type of basis spaces we have selected, will encounter the divergences in a more subtle fashion as cutoffs are elevated. For illustration, we presented an initial example in QED, without renormalization, of the mass spectrum for a single electron in a transverse cavity coupled to single photon modes.  The computational requirements of the BLFQ approach are substantial, and we foresee extensive use of leadership-class computers to obtain practical results.

\section{Acknowledgments}

The authors thank V. Karmanov, J. Hiller, J.W. Qiu and K. Tuchin for fruitful discussions.  This work was supported in part by a DOE Grant DE-FG02-87ER40371 and by DOE Contract DE-AC02-76SF00515.

\end{document}